\documentclass[14pt]{article}
\pdfoutput=1
\usepackage{amsmath,amssymb,amsfonts,amsthm,bm,bbm,cancel,wasysym}
\usepackage{epsfig,graphics,graphicx,epstopdf}
\usepackage{array,booktabs,colortbl,colordvi,multirow}
\usepackage{colordvi,color,xcolor}
\usepackage{hyperref}
\usepackage{rotating}
\usepackage{appendix}
\usepackage[utf8]{inputenc}
\usepackage{amsmath,amssymb,amsfonts,amsthm,bm,bbm,cancel,wasysym}
\usepackage{epsfig,graphics,graphicx,epstopdf}
\usepackage{array,booktabs,colortbl,colordvi,multirow}
\usepackage{colordvi,color,xcolor}
\usepackage{hyperref}
\usepackage{rotating}
\usepackage{cite}

\newcommand{\emailto}[1]{E-mail: \href{mailto:#1}{\protect \nolinkurl{#1}}} 

\begin{document}

\begin{flushright}
SLAC-PUB-17490\\
\today\\
%DRAFT\\
\end{flushright}
\vspace*{5mm}

\renewcommand{\thefootnote}{\fnsymbol{footnote}}
\setcounter{footnote}{1}

\begin{center}

{\Large {\bf Dark Matter Freeze Out With Tsallis Statistics}\\
{\bf in the Early Universe}}\\

\vspace*{0.75cm}
{\bf Thomas D. Rueter${}^{1,2}$}~\footnote{\emailto{tdr38@stanford.edu}}
{\bf Thomas G. Rizzo${}^2$}~\footnote{\emailto{rizzo@slac.stanford.edu}} and
{\bf JoAnne L. Hewett${}^2$}~\footnote{\emailto{hewett@slac.stanford.edu}},

\vspace{0.5cm}

${}^1${Stanford University, Stanford, CA, USA}

${}^2${SLAC National Accelerator Laboratory, Stanford University, Menlo Park, CA, USA}

\end{center}
\vspace{.5cm}

%-------------------------------------------------------------------------------------------------------------------------------------------------------------------

\begin{abstract}
 
\noindent
The nature of dark matter (DM) and how it might interact with the particles of the Standard Model (SM) is one of greatest mysteries currently facing particle physics, and addressing these issues should provide some understanding of how the observed relic abundance was produced. One widely considered production mechanism, a weakly interacting massive particle (WIMP) produced as a thermal relic, provides a target cross section for DM annihilation into SM particles by solving the Boltzmann equation. In this thermal freeze-out mechanism, dark matter is produced in thermal equilibrium with the SM in the early universe, and drops out of equilibrium to its observed abundance as the universe cools and expands. In this paper, we study the impact of a generalized thermodynamics, known as Tsallis statistics and governed by a parameter $q$, on the target DM annihilation cross section. We derive the phase space distributions of particles in this generalized statistical framework, and check their thermodynamic consistency, as well as analyzing the impact of this generalization on the collisional term of the Boltzmann equation. We consider the case of an initial value of $q_0>1$, with $q$ relaxing to 1 as the universe expands and cools, and solve the generalized Boltzmann numerically for several benchmark DM masses, finding the corresponding target annihilation cross sections as a function of $q_0$. We find that as $q$ departs from the standard thermodynamic case of $q=1$, the collisional term falls less slowly as a function of $x = m_\chi/T$ than expected in the standard case. We also find that the target cross section falls sharply from $\sigma v \simeq 2.2-2.6\times10^{-26} \textrm{cm}^3/\textrm{s}$ for $q_0=1$ to, for example, $\sigma v \simeq  3\times 10^{-34}  \textrm{cm}^3/\textrm{s}$ for $q_0=1.05$ for a 100 GeV WIMP.

\end{abstract}

\renewcommand{\thefootnote}{\arabic{footnote}}
\setcounter{footnote}{0}
\thispagestyle{empty}
\vfill
\newpage
\setcounter{page}{1}

%-------------------------------------------------------------------------------------------------------------------------------------------------------------------

\section{Introduction}

Current cosmological models and measurements provide compelling evidence for the existence of dark matter (DM), with an observed abundance $\Omega_\chi h^2 \simeq 0.12$ \cite{aghanim2018planck}. The DM does not interact electromagnetically, must be stable or metastable on cosmological time scales, and models of particle DM typically have interactions with Standard Model (SM) matter which provide avenues for DM production in the early universe. Recent theoretical studies have focused on mechanisms by which DM can be produced in the early universe with the observed abundance, and a wide range of models have been proposed in recent years \cite{Arcadi:2017kky, hall2010freeze, hochberg2014mechanism, kuflik2016elastically}. One of the leading candidates, thermal relic DM, proposes that the DM may `freeze-out' after being produced in thermal equilibrium with the SM in the early universe. Denoting the DM number density as $n_\chi$, freeze-out occurs as the DM-SM interaction rate $n_\chi \sigma v$ drops below the Hubble expansion rate, so that the DM interactions with the SM bath become negligible and the DM abundance is fixed. One theoretically attractive aspect of this scenario is that DM which annihilates to the SM with an interaction strength of roughly the SM weak interaction, with a thermally averaged cross section of $\left< \sigma v \right> \simeq 3\times 10^{-26} \textrm{cm}^3$/s, can generically produce the observed DM abundance for a wide range of masses $m_{\chi} \sim 10 \textrm{ GeV} - 100 \textrm{ TeV}$\cite{Arcadi:2017kky}. Such DM candidates are known as Weakly Interacting Massive Particles (WIMPs).

The thermal WIMP scenario outlined above provides one possible framework for particle theories of DM - producing the correct relic abundance of DM within a model of physics beyond the Standard Model is primarily a function of obtaining the correct value of $\left< \sigma v \right>$. However, the target value of $\left< \sigma v \right>$ mentioned above depends implicitly on the thermal history of the early universe, and modifications to that history could impact the target cross section values. In this paper we explore the impact of an extension to the standard thermodynamic picture of the early universe, through a generalized statistical mechanics first proposed by Tsallis \cite{tsallis1988, tsallis1998, tsallis2009}. 

These statistics generalize the usual Boltzmann thermodynamics via a parameter $q$ which changes the high energy behavior of the statistical distributions from the usual exponential distribution to a distribution with power-law tails, with the case of $q=1$ recovering the usual thermodynamical picture. Such power-law tails may arise due to to various high energy effects such as strong interactions or small scale fluctuations \cite{alberico2009non, biro2009non, kodama2009dyn, wilk2009power, beck2009super}, and Tsallis statistics have been used successfully to fit transverse momenta distributions at the Relativistic Heavy Ion Collider and the Large Hadron Collider \cite{abelev2007strange, adare2011identified, aamodt2011production, aad2011charged, khachatryan2011strange} with values of $q\simeq 1.1$ \cite{cleymans2012rel}. In the early universe we may expect that the quark-gluon plasma \cite{gyulassy2005new, shuryak2005rhic} exhibits similar power-law tail properties in its statistical distributions, so that the standard thermodynamical picture explored in the usual thermal freeze-out mechanism requires some generalization. The Tsallis generalization then provides a useful framework for considering deviations from the standard thermodynamics.

In the context of DM freeze-out and early universe cosmology, the case of $q$ fixed and very close to 1 has been previously studied \cite{pessah2001statistical}; here we consider larger deviations from the $q=1$ case than were previously explored in the DM context. In this paper we consider a toy model where $q=q_0\neq 1$ in the early universe while the DM is part of the thermal bath and $q$ relaxes to 1 after freeze-out occurs, in order to study the impact on the target value of $\left< \sigma v \right>$ which produces the observed DM abundance. In this toy model, we will consider initial values of $q_0 \in [1,1.05]$, and remain agnostic to the source of deviation from $q=1$ as well as the source of $q$ evolution back to 1.

In Section 2 of this paper we provide a background discussion on Tsallis statistics and thermodynamic consistency, in Section 3 we detail the impact of our generalized thermodynamics on the Boltzmann equation, in Section 4 we discuss the impacts of these generalizations on the thermal WIMP scenario, and in Section 5 we conclude.

\section{Tsallis Statistics} \label{stat-sec}

We now briefly describe the role of thermodynamics in the thermal WIMP freeze-out scenario, before deriving the necessary ingredients in the Tsallis framework. In order to evolve the Boltzmann equation as a function of $x \equiv m_\chi / T$, it is necessary to determine the Hubble parameter $H = 8 \pi G \rho /3$ as a function of the temperature $T$. Describing the energy density $\rho$ in terms of the SM particle species requires knowledge of their phase space distributions, which in the $q=1$ case are the familiar Bose-Einstein or Fermi-Dirac distributions which are functions of $E/T$. In this section, we will find phase space distributions which are functions of $q$ as well as $E/T$ within the framework of Tsallis statistics, and check that they are thermodynamically consistent, so that we can describe $\rho$ in terms of the effective degrees of freedom in a consistent way. In Section \ref{boltz-sec}, we analyze the collisional term in this generalized framework and see that it must also be modified from the familiar $q=1$ case, where it may be written as the product of phase space distributions of particle species, {\it i.e.} the assumption of molecular chaos $C[f] \sim \int \sigma v f_1 f_2$.

Tsallis statistics were proposed as a generalization of the well known Boltzmann-Gibbs statistics, based upon a generalization of the Boltzmann-Gibbs entropy, which is parameterized by a real number $q \in [0,2]$ \cite{tsallis1988}:

\begin{equation}
    S_q = k_B \frac{1-\sum_i p_i^q}{q-1} = -k_B \sum_i p_i^q \textrm{ln}_q(p_i),
\end{equation}

\noindent where $k_B$ is the Boltzmann constant, and $p_i$ represents the probability of a microstate $i$. Here we have defined a generalized logarithm, the $q$-logarithm, as $\textrm{ln}_q(x) \equiv (x^{1-q}-1)/(1-q)$, and used the normalization of the probabilities $\sum_i p_i =1$ in the second equality. In the limit $q \rightarrow 1$, $\textrm{ln}_q(x) \rightarrow \textrm{ln}(x)$ and we recover the familiar Boltzmann-Gibbs entropy $S_1 = - k_B \sum_i p_i \textrm{ln}(p_i)$. 

There is more than one version of the Tsallis phase space distributions in the literature \cite{tsallis1998, biro2009non, wilk2009power,teweldeberhan2005cut, parvan2019hadron}, so we ensure that we find a thermodynamically consistent case by maximizing a generalized entropy. We define a generalization of the entropy density four current of a particle species in terms of phase space distribution functions $f(\bf{p})$,  as \cite{cleymans2012rel,biro2012fluid} (taking $\hbar=k_B=1$):

\begin{equation} \label{ST}
    s^\mu = - g \int \frac{d^3 p}{(2 \pi)^3 p^0} ~~ p^\mu \left\{f^q \textrm{ln}_q(f) + \frac{1}{z}(1-z f)^q \textrm{ln}_q(1-z f) \right\}.
\end{equation}

\noindent Here $z = 1(-1)$ for Fermi-Dirac (Bose-Einstein) statistics, while the limit $z \rightarrow 0$ produces Maxwell-Boltzmann statistics, and $g$ represents the number of spin degrees of freedom of the particle species. Now we proceed to extremize $S^0 = V s^0$ in order to find the functional forms of $f(p)$. We use the method of Lagrange multipliers to enforce the constraints on the total energy of the system and the particle number, following the unnormalized Tsallis statistics convention \cite{cleymans2012rel}:

\begin{equation}
    g \int \frac{d^3 p}{(2 \pi)^3} f^q = \frac{N}{V} = n,
\end{equation}

\begin{equation} \label{energy}
    g \int \frac{d^3 p}{(2 \pi)^3} f^q \sqrt{m^2 + p^2} = \frac{E}{V} = \epsilon.
\end{equation}

In equilibrium we wish to maximize the entropy $s^0$ of Eq. \ref{ST} subject to constraints on the number of particles in the system $N$ and the energy of the system $E$, which leads to

\begin{equation} \label{var}
    \frac{\delta}{\delta f} \left\{ V s^0 + \alpha \left(N - g V \int \frac{d^3 p}{(2 \pi)^3} f^q \right) + \beta \left(E - g V \int \frac{d^3 p}{(2 \pi)^3} f^q \sqrt{m^2 + p^2}\right) \right\} = 0.
\end{equation}

\noindent Dividing by the common factor of $gV$, term by term, we find:

$$\frac{\delta}{\delta f} s^0 = \int \frac{d^3 p}{(2 \pi)^3} \frac{q}{q-1} \left[ \left( \frac{1-z f}{f} \right)^{q-1} -1 \right] f^{q-1},$$

$$\frac{\delta}{\delta f} \alpha \left(N - \int \frac{d^3 p}{(2 \pi)^3} f^q \right) = -\alpha \int \frac{d^3 p}{(2 \pi)^3} q f^{q-1},$$

$$\frac{\delta}{\delta f} \beta \left(E - \int \frac{d^3 p}{(2 \pi)^3} f^q \sqrt{m^2 + p^2} \right) = -\beta \int \frac{d^3 p}{(2 \pi)^3} q f^{q-1} \sqrt{m^2 + p^2}.$$

\begin{figure}
\hspace{-0.4cm}
\centerline{\includegraphics[width=3.2in,angle=0]{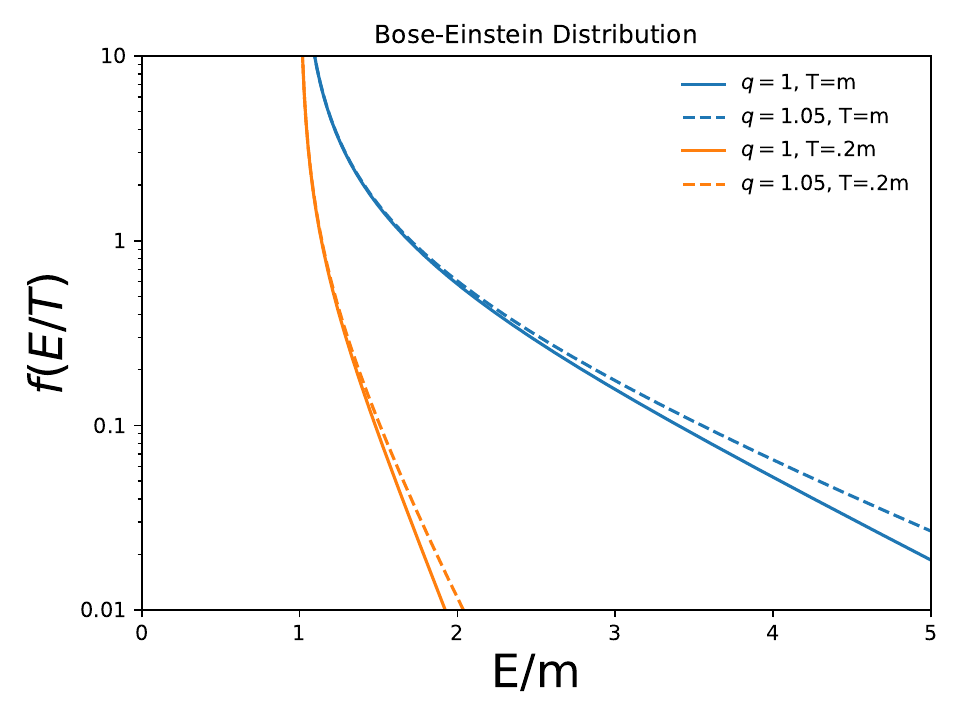}
\includegraphics[width=3.2in,angle=0]{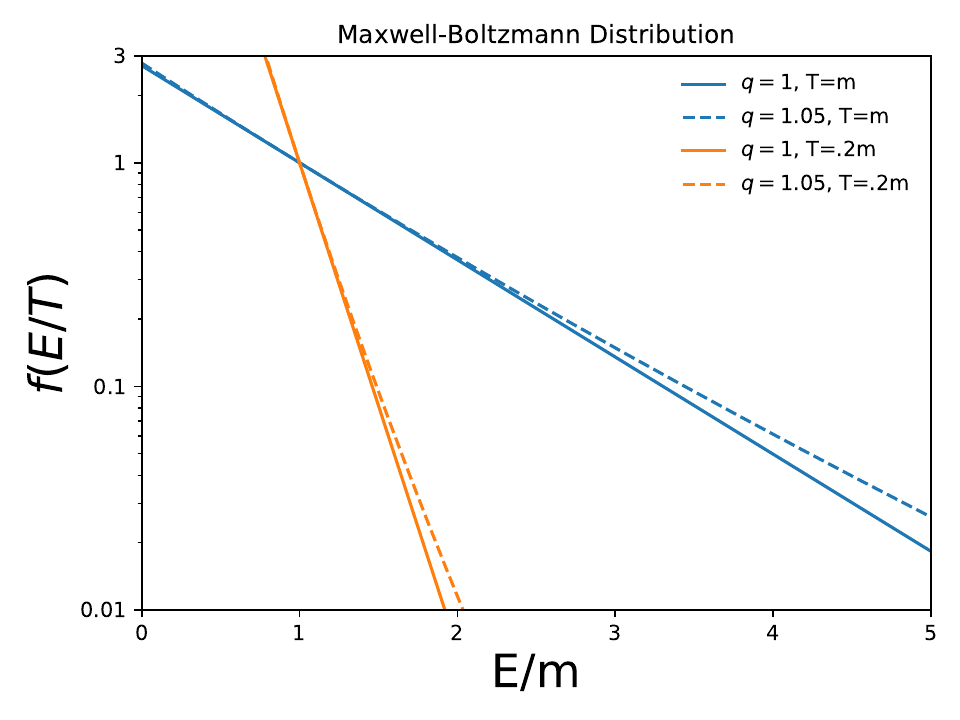}}
\centerline{\includegraphics[width=3.2in,angle=0]{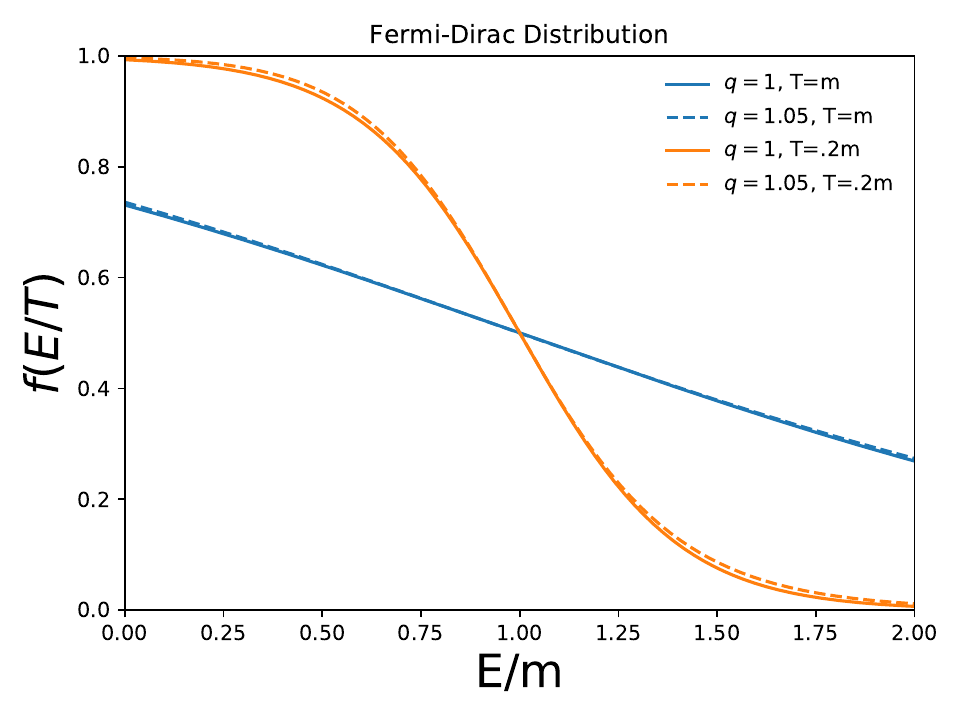}
\includegraphics[width=3.2in,angle=0]{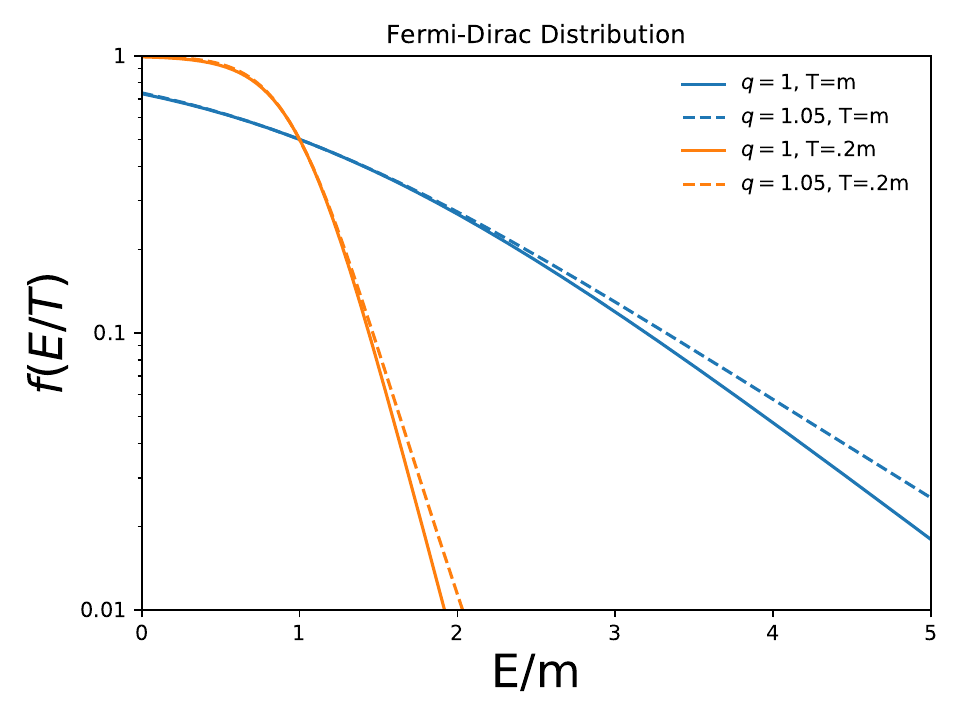}}
\caption{Equilibrium phase space distributions $f$ for Bose-Einstein (top left), Maxwell-Boltzmann (top right), and Fermi-Dirac (bottom) statistics as a function of $E/m$, with $\mu=m$. We show the distributions for $q=1$ (solid) and $q=1.05$ (dashed), taking $T=m$ (blue) and $T= m/5$ (orange) for the sake of illustration. Note that $q>1$ enhances the high energy tails of all of the distributions, and slightly enhances the low energy distributions for MB and FD statistics.} \label{f_dist}
\end{figure}

\noindent By combining these terms in Eq. \ref{var}, we find the phase space distributions to be

\begin{equation} \label{fq}
    f = \frac{1}{ [1 + (1-q)(-\beta \sqrt{m^2 +p^2} -\alpha)]^{\frac{1}{q-1}}+z} = \left[ \frac{1}{e_q(-\frac{E-\mu}{T})}+z\right]^{-1},
\end{equation}

\noindent where in the second equality we have identified $1/\beta = T$ as the temperature and $-\alpha/\beta = \mu$ as the chemical potential, and defined the $q$-exponential $e_q(x) \equiv [1+(1-q)x]^{\frac{1}{1-q}}$. The $q$-exponential is the inverse of the $q$-logarithm defined previously, $e_q(\textrm{ln}_q(x)) = \textrm{ln}_q(e_q(x)) = x$, and in the limit $q\rightarrow 1$, $e_q(x) \rightarrow e^x$. Fig. \ref{f_dist} shows $f$ for Bose-Einstein, Maxwell-Boltzmann, and Fermi-Dirac statistics at $T=m$ and $T=m/5$ for both $q=1$ and $q=1.05$. The enhancement of the high energy tails for $q>1$ is clear, and has the effect of enhancing cross sections which may otherwise be suppressed by factors of the center of mass energy. For $E < \mu$ the $f$ distributions are also slightly enhanced at $q=1.05$ relative to the $q=1$ case, but less dramatically than at high energy.

There is some ambiguity in how to treat the $q$-exponential when $1 + (1-q)x$ becomes close to zero. For $q<1$, $e_q(x)\rightarrow0$ as $x\rightarrow \frac{-1}{1-q}$, while for the case of $q>1$, $e_q(x) \rightarrow \infty$ as $x\rightarrow \frac{1}{q-1}$. In the case of $q<1$ we may define $e_{q<1}(x) \equiv 0$ for $x \leq \frac{-1}{1-q}$, which may be interpreted as a cutoff at high energies $E \geq \mu + \frac{T}{1-q}$. For $q>1$ various prescriptions have been proposed in the literature, including a ``cut-off" prescription and a modification of the $q$-exponential \cite{tsallis1988, curado1991generalized, teweldeberhan2005cut}. In the ``cut-off" approach one takes $e_q(x) \equiv 0$ for $x \geq \frac{1}{q-1}$, which would limit the number density of fermions at a given temperature for any chemical potential $\mu$, as $f \rightarrow 0$ for $E-\mu \leq \frac{-T}{q-1}$, thus cutting off $f$ at low energies. This implies a maximum number density for fermions of a given temperature for any $\mu$ above the critical value $\frac{T}{q-1}$. One method of avoiding this issue, which we include for completeness but do not adopt for our analysis, is the definition of an alternate entropy measure, as in ref \cite{teweldeberhan2005cut}, where the entropy density is defined as (suppressing factors of $2\pi$, etc) $S \sim \int d^3p ~ C_q(f)$, with $C_q(f)$ given by

\begin{equation}
C_q(f) = \left\{
	\begin{array}{ll}
		\left( \frac{f - f^q}{q-1} \right) + \left( \frac{(1-f) - (1-f)^q}{q-1} \right)  & \mbox{if } f \leq 1/2 \\
		\left( \frac{f - f^{2-q}}{1-q} \right) + \left( \frac{(1-f) - (1-f)^{2-q}}{1-q} \right) & \mbox{if } f > 1/2
	\end{array}
	\right.
\end{equation}

\noindent For $f \leq 1/2$, this definition reproduces the generalized Fermi-Dirac entropy defined with $z=1$ in Eq. \ref{ST}, and it replaces $q \rightarrow 2-q$ when $f > 1/2$ (i.e. for $E-\mu<0$). Performing the same maximization procedure as above, this alternate entropy leads to an alternate version of the phase space distribution, based upon a modified $q$-exponential:

\begin{equation}
    f_{C} = \left[\tilde e_q \left(\frac{E-\mu}{T} \right)  +1 \right]^{-1},
\end{equation}

\begin{equation}
    \tilde e_q \left( \frac{E-\mu}{T} \right) = \left\{
	\begin{array}{ll}
		\left[ 1 + (1-q)(-\frac{E-\mu}{T}) \right]^{-\frac{1}{1-q}} = \frac{1}{e_q(-\frac{E-\mu}{T})}  & \mbox{if } \frac{E-\mu}{T} > 0 \\
		\left[ 1 + (1-q)(\frac{E-\mu}{T}) \right]^{\frac{1}{1-q}} = e_q \left( \frac{E-\mu}{T} \right) & \mbox{if } \frac{E-\mu}{T} \leq 0
	\end{array}
	\right.
\end{equation}

This modified $q$-exponential is seen to be continuous and avoids the divergence at $x = -1/(1-q)$, but requires an entropy which is discontinuous at $f=1/2$: $C_q(f\rightarrow 1/2_-) = (1-2^{1-q})/(q-1)$, while $C_q(f\rightarrow 1/2_+) = 2^{q-1}(1-2^{1-q})/(q-1)$. The derivative of the entropy is continuous, however, and as $q\rightarrow1$ we recover $S = -\int d^3 p f \textrm{ln}(f)+(1-f)\textrm{ln}(1-f)$. While this alternate definition of entropy may be thermodynamically consistent, for the rest of this study we shall restrict ourselves to the continuous entropy defined in Eq. \ref{ST}, leading to the phase space distributions defined in Eq. \ref{fq}, while taking the extended-value definition of $e_q(x > \frac{1}{q-1}) \equiv \infty$. This definition implies that for values of $\frac{E-\mu}{T} < \frac{1}{1-q}$ we have $f=1$ in the Fermi-Dirac case, without the low energy cutoff behavior described above.

In order to demonstrate thermodynamic consistency, several thermodynamic derivative relations from the first and second laws must hold between the energy density $\epsilon \equiv E/V$, the entropy density $s \equiv S/V$, the number density $n \equiv N/V$, and the pressure $P$ \cite{reif2009}: 

\begin{equation} \label{T1}
    \frac{\partial \epsilon}{\partial s} \bigg\rvert_n = T,
\end{equation}
\begin{equation} \label{T2}
    \frac{\partial P}{\partial \mu} \bigg\rvert_T = n,
\end{equation}
\begin{equation} \label{T3}
     \frac{\partial P}{\partial T} \bigg\rvert_\mu = s,
\end{equation}
\begin{equation} \label{T4}
    \frac{\partial \epsilon}{\partial n} \bigg\rvert_s = \mu.
\end{equation}

Eqs. \ref{T1} and \ref{T2} were demonstrated to be valid in ref \cite{cleymans2012rel} for the equilibrium distributions as defined in Eq. \ref{fq}. Before explicitly calculating Eqs. \ref{T3} and \ref{T4}, we demonstrate that the first law of thermodynamics holds with the entropy defined in Eq. \ref{ST} for the equilibrium distributions of Eq. \ref{fq}. Beginning with the entropy density $s$

\begin{equation} \label{s}
    s = - g \int \frac{d^3 p}{(2 \pi)^3}  \left\{f^q \textrm{ln}_q(f) + \frac{1}{z}(1-z f)^q \textrm{ln}_q(1-z f) \right\},
\end{equation}

\noindent we wish to show that $Ts = \epsilon + P -\mu n$. Adding and subtracting $\epsilon/T - \mu n /T$ on the right hand side of Eq. \ref{s}, we see that the first law of thermodynamics requires that

\begin{equation} \label{P1}
    \frac{P}{T} = -g \int \frac{d^3 p}{(2 \pi)^3}  (f^q \textrm{ln}_q(f) + \frac{1}{z}(1-z f)^q \textrm{ln}_q(1-z f) + \frac{E}{T} f^q - \frac{\mu}{T} f^q. 
\end{equation}

\noindent The integrand can be simplified employing the relations

$$1 - z f = \frac{f}{e_q(-\frac{E-\mu}{T})},$$
$$\left[e_q(x)\right]^q = \frac{e_q(x)}{1+(1-q)x},$$

\noindent where we have used the equilibrium distributions of Eq. \ref{fq} and the definition of $e_q(x)$. The integral then becomes

\begin{equation}
    -g \int \frac{dE}{2 \pi^2} E \sqrt{E^2 - m^2} \frac{1}{z} f^{q-1} \left(\frac{E-\mu}{T} + \textrm{ln}_q(f)\right) = g \int \frac{d^3 p}{(2 \pi)^3} \frac{p^2}{3E} \frac{f^q}{T} = \frac{P}{T},
\end{equation}

\noindent where we have integrated by parts in the first equality, and used the standard definition of pressure in terms of the equilibrium mean occupancy number in the second equality \cite{baumann2012cosmology}. We can now proceed similarly to ref \cite{cleymans2012rel} in demonstrating the validity of the thermodynamic relations \ref{T3} and \ref{T4}. Using $P = -\epsilon + Ts +\mu n$ we write

\begin{equation}
    \begin{aligned}
    \frac{\partial P}{\partial T} \bigg\rvert_\mu &= -\frac{\partial \epsilon}{\partial T} + s + T \frac{\partial s}{\partial T} + \mu \frac{\partial n}{\partial T} \\
    &= s + g\int \frac{d^3 p}{(2 \pi)^3} \left[ -q E + T \frac{q}{q-1} \left\{ \left( \frac{1-zf}{f} \right)^{q-1} -1 \right\} + q\mu \right] f^{q-1} \frac{\partial f}{\partial T} \\
    &= s,
    \end{aligned}
\end{equation}

\noindent where we have used Eq. \ref{var} with $\beta = 1/T$ and $\alpha = -\mu/T$ to see that the integrand in the second line is identically zero for the equilibrium distributions, and thus that Eq. \ref{T3} is satisfied. 

Eq. \ref{T4} can be evaluated as 

\begin{equation} \label{dedn1}
    \frac{\partial \epsilon}{\partial n} \bigg\rvert_s = \frac{\frac{\partial \epsilon}{\partial T} |_s dT + \frac{\partial \epsilon}{\partial \mu} |_s d\mu}{\frac{\partial n}{\partial T} |_s dT + \frac{\partial n}{\partial \mu} |_s d\mu} = \frac{\frac{\partial \epsilon}{\partial T} |_s + \frac{\partial \epsilon}{\partial \mu} |_s \frac{d\mu}{dT}}{\frac{\partial n}{\partial T} |_s + \frac{\partial n}{\partial \mu} |_s \frac{d\mu}{dT}}.
\end{equation}

\noindent Since this is evaluated holding $s$ constant, $ds = (\partial s/\partial \mu) d\mu + (\partial s/\partial T) dT =0$. This then implies 

\begin{equation} \label{dmdT}
    \frac{d \mu}{dT} = -\frac{\frac{\partial s}{\partial T}}{\frac{\partial s}{\partial \mu}} = - \frac{\int \frac{d^3 p}{(2 \pi)^3} \left( \frac{E-\mu}{T} \right)^2 f^{q+1}\left[1+(1-q)(-\frac{E-\mu}{T}) \right]^{-\frac{1}{1-q}-1}}{\int \frac{d^3 p}{(2 \pi)^3} \left( \frac{E-\mu}{T} \right) f^{q+1}\left[1+(1-q)(-\frac{E-\mu}{T}) \right]^{-\frac{1}{1-q}-1}},
\end{equation}

\noindent so that Eq. \ref{dedn1} may be written

\begin{equation} \label{dedn2}
     \frac{\partial \epsilon}{\partial n} \bigg\rvert_s = \frac{\int \frac{d^3 p}{(2 \pi)^3} \frac{qE}{T}f^{q+1} \left[1+(1-q)(-\frac{E-\mu}{T}) \right]^{-\frac{1}{1-q}-1} \left(\frac{E-\mu}{T}  + \frac{d \mu}{dT}  \right)}{\int \frac{d^3 p}{(2 \pi)^3} \frac{q}{T}f^{q+1} \left[1+(1-q)(-\frac{E-\mu}{T}) \right]^{-\frac{1}{1-q}-1} \left(\frac{E-\mu}{T}  + \frac{d \mu}{dT}  \right)}.
\end{equation}

\noindent Adding and subtracting $\mu$ on the right hand side of \ref{dedn2} we may write:

\begin{equation}
    \begin{aligned}
        \frac{\partial \epsilon}{\partial n} \bigg\rvert_s &= \mu + \frac{\int \frac{d^3 p}{(2 \pi)^3} \frac{q(E-\mu)}{T}f^{q+1} \left[1+(1-q)(-\frac{E-\mu}{T}) \right]^{-\frac{1}{1-q}-1} \left(\frac{E-\mu}{T}  + \frac{d \mu}{dT}  \right)}{\int \frac{d^3 p}{(2 \pi)^3} \frac{q}{T}f^{q+1} \left[1+(1-q)(-\frac{E-\mu}{T}) \right]^{-\frac{1}{1-q}-1} \left(\frac{E-\mu}{T}  + \frac{d \mu}{dT}  \right)} \\
        &=\mu + \frac{\int \frac{d^3 p}{(2 \pi)^3} q (\frac{E-\mu}{T})^2 G(\frac{E-\mu}{T})  + \frac{d \mu}{dT}\int \frac{d^3 p}{(2 \pi)^3} q (\frac{E-\mu}{T})G(\frac{E-\mu}{T})}{\int \frac{d^3 p}{(2 \pi)^3} \frac{q}{T} G(\frac{E-\mu}{T}) \left(\frac{E-\mu}{T}  + \frac{d \mu}{dT}  \right)} \\
        &= \mu,
    \end{aligned}
\end{equation}

\noindent so that the relationship of Eq. \ref{T4} is satisfied. In the second equality we have defined $G(\frac{E-\mu}{T}) \equiv f(\frac{E-\mu}{T})^{q+1} \left[1+(1-q)(-\frac{E-\mu}{T}) \right]^{-\frac{1}{1-q}-1}$, and in the third equality we have used Eq. \ref{dmdT}. Having shown that the equilibrium distributions found by maximizing the entropy of Eq. \ref{ST} are thermodynamically consistent, we now examine how this thermodynamic generalization impacts the Boltzmann equation. 

\section{Generalized Boltzmann Equation} \label{boltz-sec}

The usual Boltzmann equation can be written as 

\begin{equation}
    L[f] = C[f],
\end{equation}

\noindent where $L[f]$ is the Liouville operator, and $C[f]$ is the collision integral. Lavagno  \cite{lavagno2002rel} first attempted to generalize this equation in the context of the thermodynamics outlined in Section \ref{stat-sec}, though with a different version of the entropy four current. Here we follow a similar procedure, with the entropy four current density as defined in Eq. \ref{ST}, and consider the entropy production rate

\begin{equation} \label{ent1prod}
    \partial_\mu S^{\mu} = - g \int \frac{d^3 p}{(2 \pi)^3 p^0} [\textrm{ln}_q(f) - f^{1-q}(1-z f)^{q-1}\textrm{ln}_q(1-zf)] p^\mu \partial_\mu(f^q).
\end{equation}

Now using the generalized Boltzmann equation as proposed by Lavagno \cite{lavagno2002rel}, $p^\mu \partial_\mu(f^q) = C[f]$, we can calculate the entropy production in a $2\rightarrow2$ scattering process. We define the collision integral as 

\begin{equation}
C[f_1] = \int dP_2 dP_3 dP_4\left\{ W_{34\rightarrow12}H_q[f_3,f_4;f_1,f_2] - W_{12\rightarrow34}H_q[f_1,f_2;f_3,f_4] \right\}
\end{equation},

\noindent where  $W_{12\rightarrow34}$ is a Lorentz-invariant scattering rate and $dP_i = g_i d^3 p_i /((2 \pi)^3 p_i^0)$, with $g_i$ the number of spin degrees of freedom of species $i$. $H_q[f_1,f_2;f_3,f_4]$ is an function which modifies the rate of the $12\rightarrow34$ reaction based on the values of the phase space distributions $f_i$, and it is this function we wish to investigate. Taking $W_{12\rightarrow34} = W_{34\rightarrow12}$, {\it i.e.} assuming the microphysics is reversible, we can calculate the total entropy production of this process as

\begin{equation}
    \begin{aligned}
    \partial_\mu S^{\mu}_{12\rightarrow34} = \int dP_1 dP_2 dP_3 dP_4 \{W_{12\rightarrow34} & (F[f_3]+F[f_4]-F[f_1]-F[f_2])\\
    & \times (H_q[f_1,f_2;f_3,f_4]-H_q[f_3,f_4;f_1,f_2])\},
    \end{aligned}
\end{equation}

\noindent where we have defined $F[f_i] = -\textrm{ln}_q(f_i) + f_i^{1-q}(1-z_i f_i)^{q-1}\textrm{ln}_q(1-z_i f_i)$ from Eq. \ref{ent1prod}. In order to have $\partial_\mu S_{12\rightarrow34}^\mu \geq 0$, we define
\begin{equation}
    \begin{aligned}
    H_q[f_1,f_2;f_3,f_4] = e_q\big[ &\textrm{ln}_q(f_1) + \textrm{ln}_q(f_2) + f_3^{1-q}(1-z_3 f_3)^{q-1}\textrm{ln}_q(1-z_3 f_3) \\ &+ f_4^{1-q}(1-z_4 f_4)^{q-1}\textrm{ln}_q(1-z_4 f_4) \big],
    \end{aligned}
\end{equation}

\noindent which is the $q$ generalized version of the assumption of molecular chaos. As $q\rightarrow1$, we recover the familiar case $H_1[f_1,f_2;f_3,f_4] = f_1 f_2 (1-z_3 f_3) (1-z_4 f_4)$, where $z_i = +1(-1)$  for fermions (bosons). When quantum statistics become unimportant, taking $z_i\rightarrow0$ yields $H_q[f_1,f_2] = e_q[\textrm{ln}_q(f_1) + \textrm{ln}_q(f_2)]$, which is the case proposed by Lavagno \cite{lavagno2002rel}. 

We now consider the Boltzmann equation, taking the Liouville operator to act on $f^q$ rather than $f$ in light of the discussion above:

\begin{equation}
    L[f^q] = E \partial_t (f^q) - H |\vec{p}|^2 \frac{\partial (f^q)}{\partial E} = C[f].
\end{equation}

Dividing by the energy and integrating $L[f^q]$ over the momenta and summing over spin degrees of freedom gives us the familiar result $\partial_t n + 3 H n = a^{-3}\partial_t (n a^3)$. The right hand side then becomes
 
\begin{align} \label{boltz1}
    g_1  \int \frac{d^3 p_1}{(2 \pi)^3 E_1} C[f] = & - \sum_{spins} \int \frac{d^3 p_1}{(2 \pi)^3 2 E_1} \int \frac{d^3 p_2}{(2 \pi)^3 2 E_2} \int \frac{d^3 p_3}{(2 \pi)^3 2 E_3} \int \frac{d^3 p_4}{(2 \pi)^3 2 E_4} \\
    & \times (2 \pi)^4 \delta^{(4)}(p_1 + p_2 - p_3 -p_4) \\
    & \times \left(  H_q[f_1, f_2;f_3,f_4] \left| \mathcal{M}_{12\rightarrow 34} \right|^2 -  H_q[f_3, f_4;f_1,f_2] \left| \mathcal{M}_{34\rightarrow 12} \right|^2 \right).
\end{align}

\begin{figure}
\hspace{-0.4cm}
\centerline{\includegraphics[width=3.2in,angle=0]{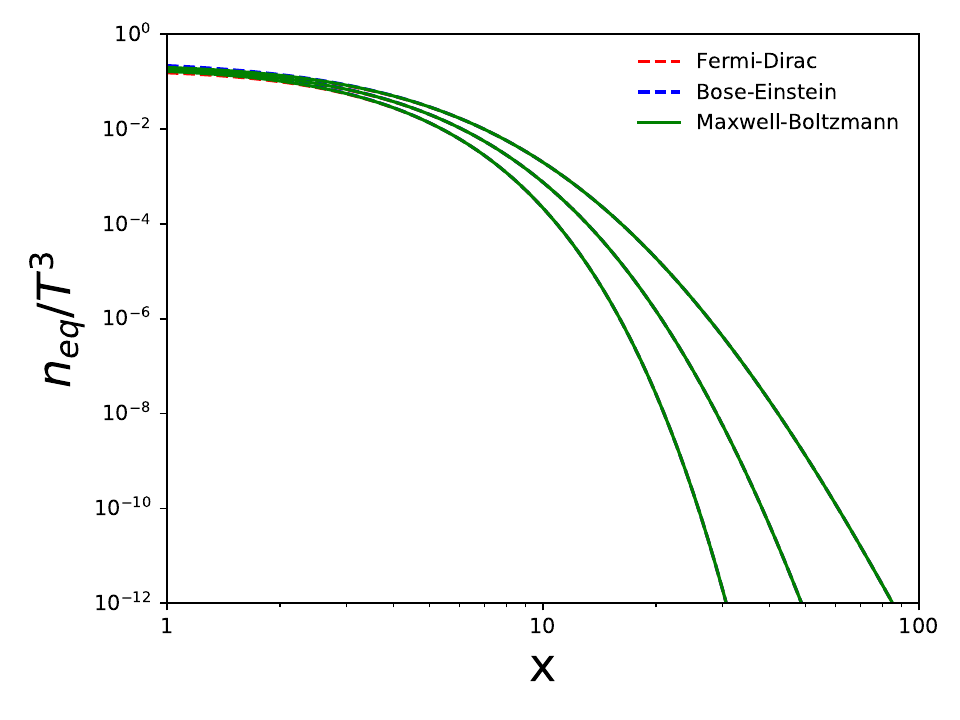}
\includegraphics[width=3.2in,angle=0]{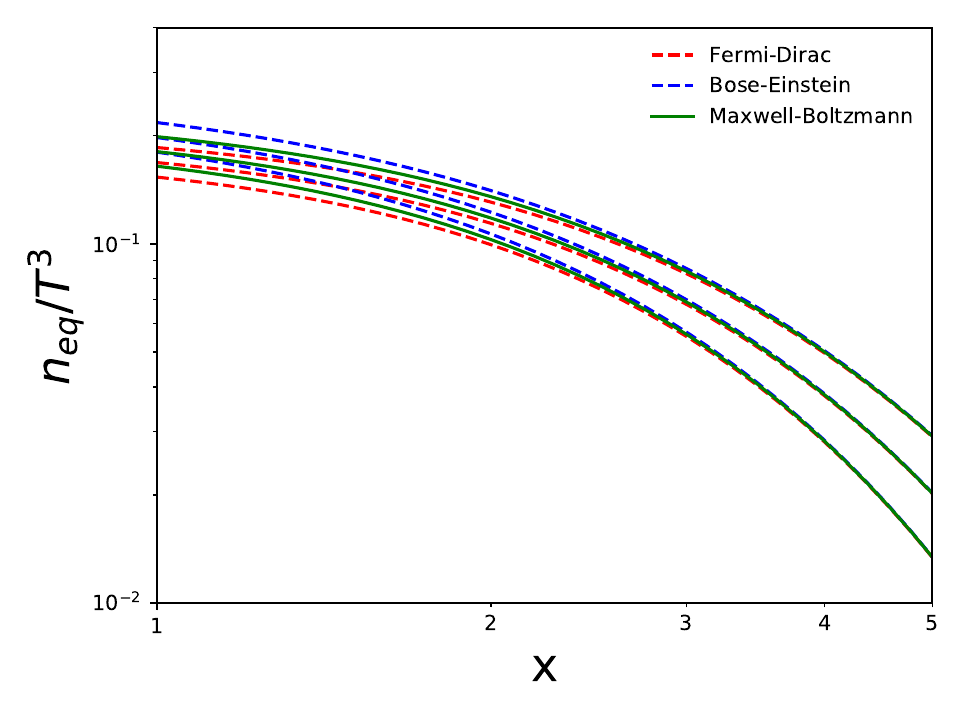}}
\caption{Left panel: Equilibrium abundances for $n_{eq}/T^3$ for Fermi-Dirac (red dashed), Maxwell-Boltzmann (green), and Bose-Einstein (blue dashed) statistics for $q$=1.05 (top), $q$=1.025 (middle), and $q$=1 (bottom), as a function of $x=m/T$, taking $g=2$. Right panel: Same as in the left panel, but zoomed in to show that quantum statistics become unimportant for $x \gtrsim 5$.} \label{neq}
\end{figure}

Since we are investigating the thermal WIMP scenario we will take particles species 1 and 2 to be DM, and 3 and 4 to be SM states, which will be in equilibrium during the freeze-out process. As can be seen in Fig. \ref{neq}, the DM may be approximated by Maxwell-Boltzmann statistics ($z=0$) for $x=m/T \gtrsim 5$, and in this regime we neglect the effects of Pauli blocking and Bose enhancement for the SM particles. The detailed balance condition in equilibrium thus becomes $H_q[f_{1,eq},f_{2,eq}] = H_q[f_{3,eq},f_{4,eq}]$, so we can replace $H_q[f_3,f_4]$ in Eq. \ref{boltz1} with $H_q[f_{1,eq},f_{2,eq}]$. Unitarity allows us to write 

\begin{align}
    &\int \frac{d^3 p_3}{(2 \pi)^3 2 E_3} \int \frac{d^3 p_4}{(2 \pi)^3 2 E_4} (2 \pi)^4 \delta^{(4)}(p_1 + p_2 - p_3 -p_4) \left| \mathcal{M}_{12\rightarrow 34} \right|^2 = \\
    & \int \frac{d^3 p_3}{(2 \pi)^3 2 E_3} \int \frac{d^3 p_4}{(2 \pi)^3 2 E_4} (2 \pi)^4 \delta^{(4)}(p_1 + p_2 - p_3 -p_4) \left| \mathcal{M}_{34\rightarrow 12} \right|^2.
\end{align}

This allows us to perform the integrals over $d^3 p_3 d^3 p_4$, which gives the unpolarized cross section $\sigma_{12\rightarrow 34}$ \cite{gondolo1991cosmic}:

\begin{align}
    & \sum_{spins} \int \frac{d^3 p_3}{(2 \pi)^3 2 E_3} \int \frac{d^3 p_4}{(2 \pi)^3 2 E_4} (2 \pi)^4 \delta^{(4)}(p_1 + p_2 - p_3 -p_4) \left| \mathcal{M}_{12\rightarrow 34} \right|^2 \\
    & = 4 g_1 g_2 \left( (p_1\cdot p_2)^2 - m_1^2 m_2^2 \right)^{1/2} \sigma_{12\rightarrow 34}.
\end{align}

We must perform a sum over all final states to get the total cross section $\sigma = \sum_f \sigma_{12 \rightarrow f}$. Taking $m_1 = m_2 = m$ as the DM mass, and using the Mandelstam variable $s=(p_1+p_2)^2$ we can thus express $C[f]$ as

\begin{equation} \label{collf}
    g_1 \int \frac{d^3 p_1}{(2 \pi)^3 E_1} C[f] = \int \frac{d^3 p_1}{(2 \pi)^3 2 E_1} \int \frac{d^3 p_2}{(2 \pi)^3 2 E_2} 4 g_1 g_2 \frac{1}{2} \sqrt{s(s-4m^2)} \sigma (H_q[f_{1,eq},f_{2,eq}] - H_q[f_1,f_2]).
\end{equation}

At this point in the $q=1$ case it is convenient to define a thermally averaged cross section by taking $f = e^{-(E-\mu)/T}$ and factoring out the dependence on the chemical potential \cite{gondolo1991cosmic}, however it is impossible to do so in the $q\neq1$ case since $e_q(x)e_q(y) = e_q(x+y+(1-q)xy)$. Instead we must calculate the collision integral numerically. We first make a change of variables to $E_+ = E_1 + E_2$, $E_- = E_1 - E_2$, and $s = (p_1 + p_2)^2$

$$d^3p_1 d^3p_2 = 8 \pi^2 p_1 E_1 dE_1 p_2 E_2 dE_2 d\textrm{cos}\theta = 2\pi^2 E_1 E_2 dE_+ dE_-ds, $$

\noindent so that the integral becomes

\begin{equation}
     \frac{2 \pi^2 g_1 g_2}{(2 \pi)^6} \int_{4m^2}^{\infty} ds \int_{\sqrt{s}}^{\infty} dE_+ \int_{-\sqrt{1-\frac{4m^2}{s}}\sqrt{E_+^2-s}}^{\sqrt{1-\frac{4m^2}{s}}\sqrt{E_+^2-s}} dE_-  \frac{1}{2}\sqrt{s(s-4m^2)} \sigma (H_q[f_{1,eq},f_{2,eq}] - H_q[f_1,f_2]).
\end{equation}

To make further progress we must use the explicit forms of $H_q$ and $f$. Taking $H_q[f_1,f_2] = e_q(\textrm{ln}_q(f_1) + \textrm{ln}_q(f_2))$, and with $f=e_q(-\frac{E-\mu}{T})$, we see that the integrand can be written in terms of $E_+$ and $s$ as:

\begin{equation}
     \frac{2 \pi^2 g^2}{(2 \pi)^6} \int_{4m^2}^{\infty} ds \int_{\sqrt{s}}^{\infty} dE_+  (s-4m^2)\sigma \sqrt{E_+-s}  \left[e_q\left(-\frac{E_+}{T} \right) - e_q\left(-\frac{E_+-2\mu}{T}\right) \right].
\end{equation}

\noindent Here we have used the fact that particles 1 and 2 are DM, so that $g_1 = g_2 = g$ and $\mu_1=\mu_2=\mu$, and performed the trivial integral over $E_-$. It will be convenient to define the collision integral with its mass dimensions factored out. To do so, we make a second change of variables

\begin{equation}
\begin{aligned}
    & \epsilon = \frac{s-4m^2}{4m^2}, \\
    & u = \frac{E_+}{\sqrt{s}},\\
\end{aligned}
\end{equation}

\noindent where the variable of integration $\epsilon$ is not to be confused with the energy density of Eq. \ref{energy} so that the collision integral becomes

\begin{equation} \label{presv}
    \frac{128 \pi^2 g^2}{(2\pi)^6}m^6\int_0^{\infty} d\epsilon  (\epsilon+1)\epsilon \int_1^{\infty} du \sqrt{u^2-1}\sigma \left[e_q(-2x \sqrt{\epsilon+1}u) - e_q(-2x(\sqrt{\epsilon+1}u - y)) \right],
\end{equation}

\noindent where we have introduced the familiar ratio $x \equiv m/T$, which will serve as the time variable in the Boltzmann equation, and $y \equiv \mu/m$. Since freeze-out happens at temperatures far below the DM mass, it is sometimes useful to expand the cross section in powers of velocity. In particular we wish to expand $\sigma v_{lab}$, where the cross section times velocity is determined in the rest frame of one of the particles, in powers of $\epsilon$, which is the kinetic energy per unit mass in this frame. For a two particle final state, $\sigma v_{lab}$ is given by \cite{gondolo1991cosmic}:

\begin{equation}
    \sigma v_{lab} = \frac{\beta_f}{64 \pi^2 (s-2m^2)}\int d\Omega \left| \mathcal{M} \right|^2,
\end{equation}

\noindent where $d\Omega = d\textrm{cos}\theta d\phi$ and $\theta$ is the center of mass scattering angle and 

\begin{equation}
    \beta_f = \sqrt{1-\frac{(m_3+m^4)^2}{s}}\sqrt{1-\frac{(m_3-m_4)^2}{s}}
\end{equation}
is an invariant quantity which corresponds to the velocity of the final state particles in the center of mass frame for $m_3=m_4$. In the lab frame we can write

\begin{equation}
    v_{lab} = \frac{\frac{1}{2}\sqrt{s(s-4m^2)}}{E_{1,lab}E_{2,lab}} = \frac{2\sqrt{\epsilon(\epsilon+1)}}{1+2\epsilon},
\end{equation}

\noindent and thus rewrite the collision integral as

\begin{equation} \label{collsv}
    \begin{aligned}
    \frac{64 \pi^2 g^2}{(2\pi)^6}m^6 &\int_0^{\infty} d\epsilon  \sqrt{\epsilon(\epsilon+1)}(1+2\epsilon) \int_1^{\infty} du \sqrt{u^2-1}\\
    & \times \sigma v_{lab} \left[e_q(-2x \sqrt{\epsilon+1}u) - e_q(-2x(\sqrt{\epsilon+1}u - y)) \right]. \\ 
    \end{aligned}
\end{equation}

\begin{figure}
\hspace{-0.4cm}
\centerline{\includegraphics[width=3.2in,angle=0]{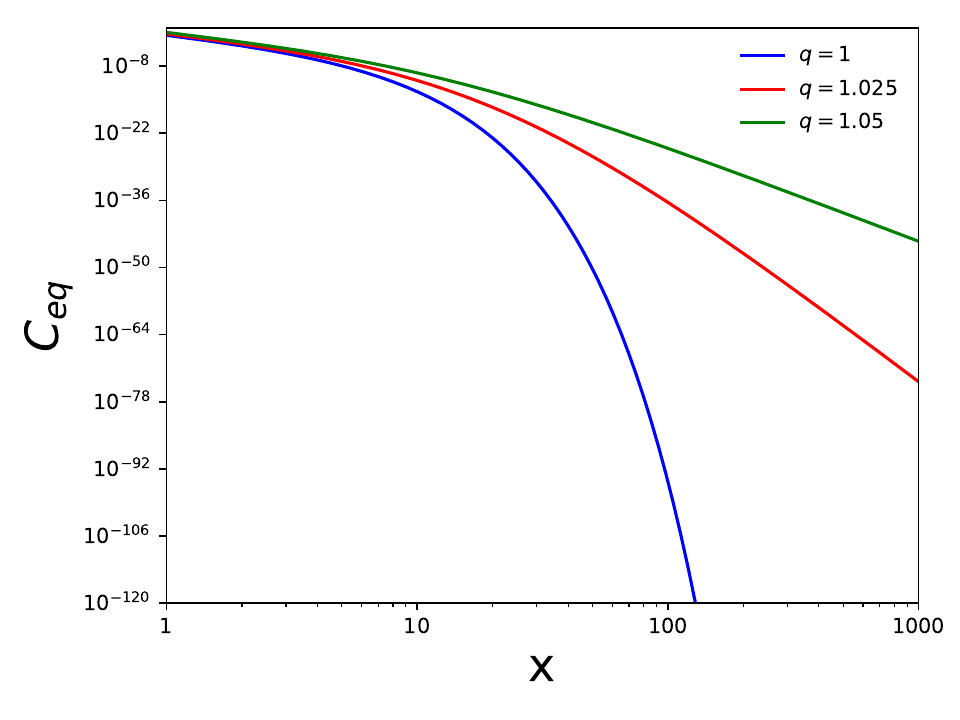}
\includegraphics[width=3.2in,angle=0]{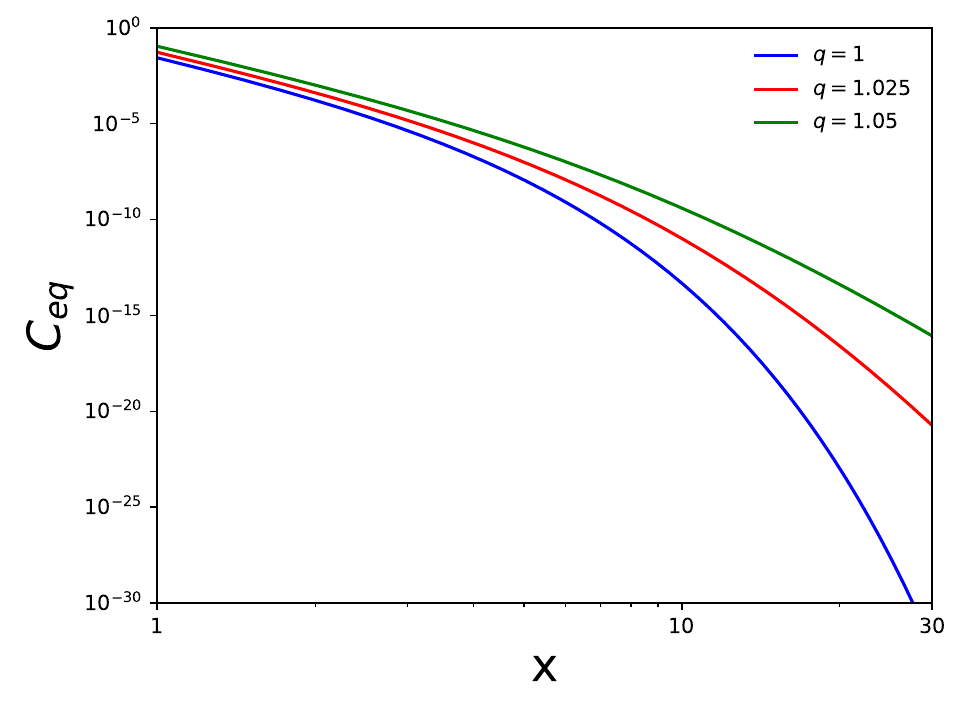}}
\caption{Left panel: The equilibrium ($y=0$) collisional term of Eq. \ref{collsv} as a function of $x=m/T$ for $q=1$ (blue), $q=1.025$ (red), and $q=1.05$ (green), apart from an overall factor of $m^6 a_0$ for the purpose of illustration. Increasing $q > 1$ results in order of magnitude enhancements relative to the $q=1$ case at large $x$ . Right panel: Same as left, zoomed in to show behavior at small $x$. At $x=1$, the enhancement for $q>1$ is only a factor of a few, but it grows exponentially as $x$ increases.} \label{heqfixq}
\end{figure}

Expanding $\sigma v_{lab} = a_0 + a_1 \epsilon + a_2 \epsilon^2+...$, we can numerically evaluate the relative strength of the leading order terms as a function of $x$. We refer to the leading terms of $a_0$, $a_1$, and $a_2$ as $s$-, $p$-, and $d$-wave annihilations, respectively. Fig. \ref{heqfixq} shows the equilibrium distribution collisional term, the first term in Eq. \ref{collsv}, as a function of $x=m/T$ for several values of $q$ in the $s$-wave case. At low $x$, increasing $q$ increases the collisional term by a factor of a few, but as $x$ increases the enhancement of $q > 1$ cases relative to $q=1$ grows exponentially so that at large $x$ the enhancement is many orders of magnitude. Even when $x=25$, an approximate value for DM freeze-out in the usual WIMP scenario, the $q=1.05$ case is approximately $6.4 \times 10^{12}$ larger than that in the $q=1$ case. This behavior causes the DM to remain in chemical equilibrium with the SM for potentially much longer when $q>1$, given the same value of the $s$-wave cross section $a_0$.

\begin{figure}
\hspace{-0.4cm}
\centerline{\includegraphics[width=3.1in,angle=0]{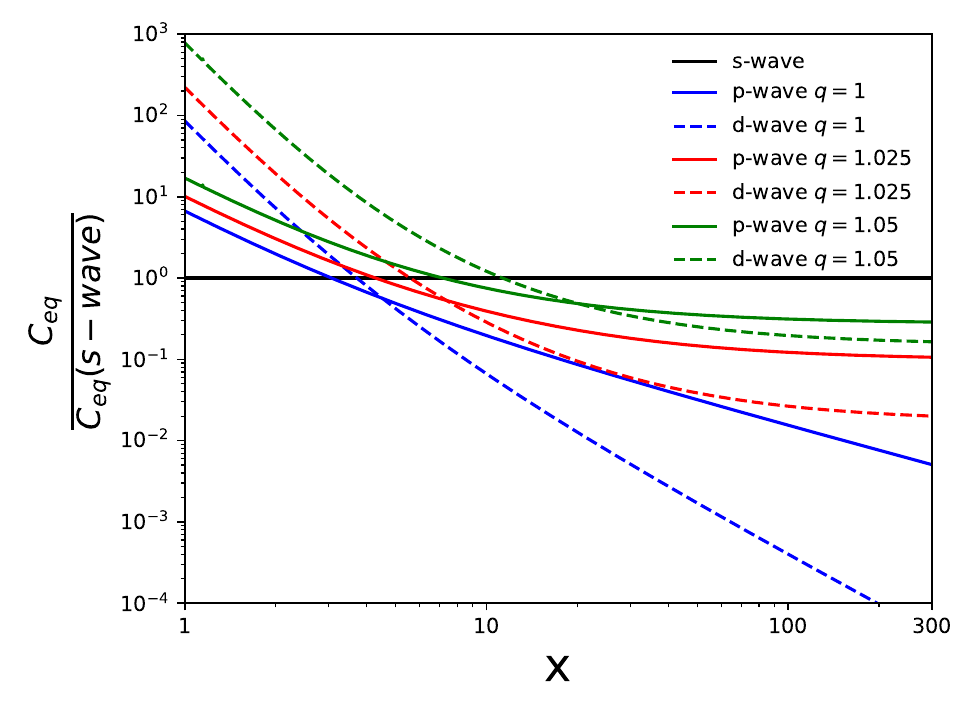}
\includegraphics[width=3.1in,angle=0]{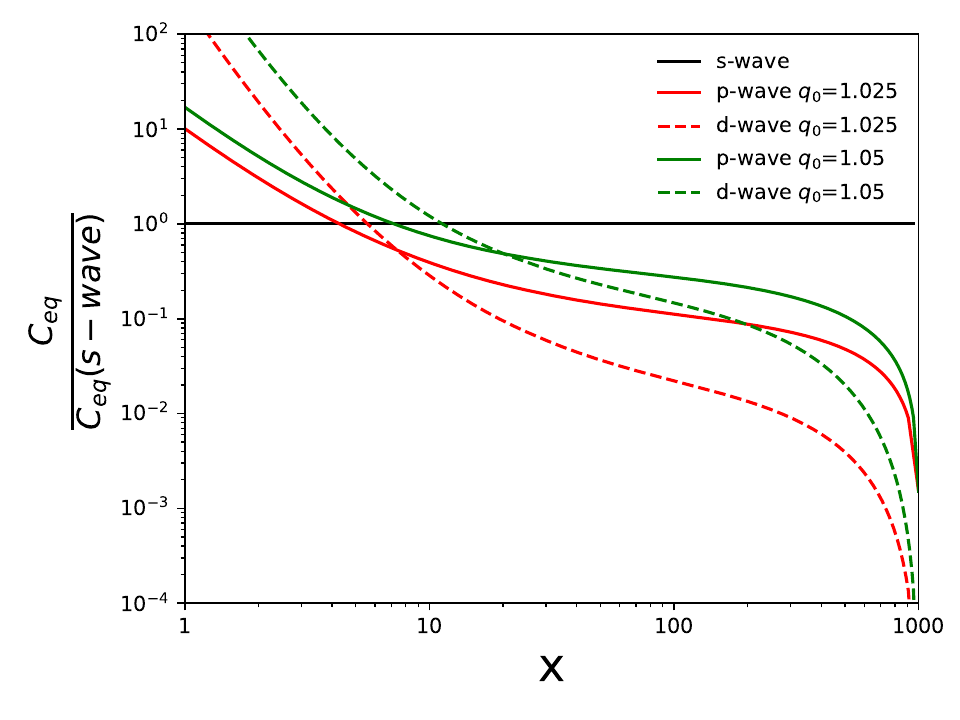}}
\caption{Left panel: Ratios of $p$-wave (solid) and $d$-wave (dashed) annihilation relative to the $s-$wave case as a function of $x=m/T$ for $q=1.05$ (green), $q = 1.025$ (red) and $q=1$ (blue). Here we have taken $a_0=a_1=a_2$ so that the differences arise solely from the leading powers of $\epsilon$ in the expansion. As $q$ deviates further from 1, the $p$- and $d$-wave annhilations are less suppressed relative to the $s$-wave case than in the $q=1$ case, and they fall much more slowly at large $x$. Right panel: The same annihilation ratios, but with $q$ evolving linearly in $x$ from 1.05 at DM freeze-out ($x_f \simeq 25$) to 1 at $x=1000$ as described in Sec. \ref{disc-sec}. The ratios drop sharply when $q$ becomes very close to 1 at large $x$.} \label{heqspd}
\end{figure}

The left panel of Fig. \ref{heqspd} shows ratios of the $p$- and $d$-wave annihilation terms to the $s$-wave case for several values of $q$. The impact of the enhanced tails of the phase space distributions and $e_q(x)$ for $q>1$ relative to $q=1$ manifests itself in the slower drop off of $p$- and $d$-wave annihilations with increasing $x$ relative to the $s$-wave case, with the suppression levelling off as $x$ approaches 1000. As $q$ increases, the impact of the long tails grows, and the $x$ value at which $p$- and $d$-wave annihilation rates drop below $s$-wave rates increases. Larger values of $q$ also reduce the magnitude of the $p$- and $d$-wave suppression factors, most notably at large $x$. For example, for $x\simeq 200$ the $p$- and $d$-wave suppressions are the same order of magnitude for $q=1.05$, whereas usually the $d$-wave case is suppressed by two orders of magnitude relative to the $p$-wave case for $q=1$. This implies that when $q>1$ we cannot make use of the usual approximations that a $p$-wave ($d$-wave) cross section can be treated as having a $1/x$ ($1/x^2$) suppression relative to the $s$-wave case for large values of $x$. The right panel of Fig. \ref{heqspd} shows the same ratios, but with $q$ evolving from 1.05 to 1 as described in Sec. \ref{disc-sec}. While both the $p$- and $d$-wave cases drop off sharply at large $x$ as $q\rightarrow1$, we see that the $d$-wave case drops by over three orders of magnitude between $x=100$ and $x=1000$ for $q_0=1.05$, while the $p$-wave case drops by approximately 2 orders of magnitude over the same range. This indicates that when $q>1$ $d$-wave annihilations are more greatly enhanced than $p$-wave annihilations relative to the $q=1$ case.

Returning to the Boltzmann equation, we divide both sides by the entropy density $s$, and then use the conservation of comoving entropy $S = a^3 s$ to write the left hand side in terms of $Y = n/s$. Writing $s = (\rho + P)/T$ for the SM in chemical equilibrium, the conservation of comoving entropy

\begin{equation}
    \frac{d S}{d t} = \frac{d}{dt} \left[ \frac{\rho + P}{T} V \right] = 0
\end{equation}

\noindent can be shown from the conservation of stress-energy $\nabla_\mu T^\mu _\nu = 0$ and the thermodynamic relation of Eq. \ref{T3}, where we have written $s = (\rho + P)/T$ in chemical equilibrium. Changing time variables to $x=m/T$ the Boltzmann equation then becomes

\begin{equation}
    \frac{dY}{dx} = - \frac{m^6}{Hxs}\left[ C_{non-eq} - C_{eq} \right],
\end{equation}

\noindent where $C_{eq}$ and $C_{non-eq}$ are the equilibrium and non-equilibrium collision integrals in Eq. \ref{collsv} with the factors of $m^6$ removed. $H$ and $s$ can be written in terms of the number of energy and entropy degrees of freedom $g_{\textrm{eff}}$ and $h_{\textrm{eff}}$, respectively. We define these in terms of the number of equivalent $q=1$ relativistic bosonic degrees of freedom, so that

\begin{equation}\label{rhos}
    \begin{aligned}
    \rho &= \frac{\pi^2}{30} g_{\textrm{eff}} T^4, \\
    s &= \frac{2 \pi^2}{45} h_{\textrm{eff}} T^3.\\
    \end{aligned}
\end{equation}

\begin{figure}
\centerline{\includegraphics[scale=0.7]{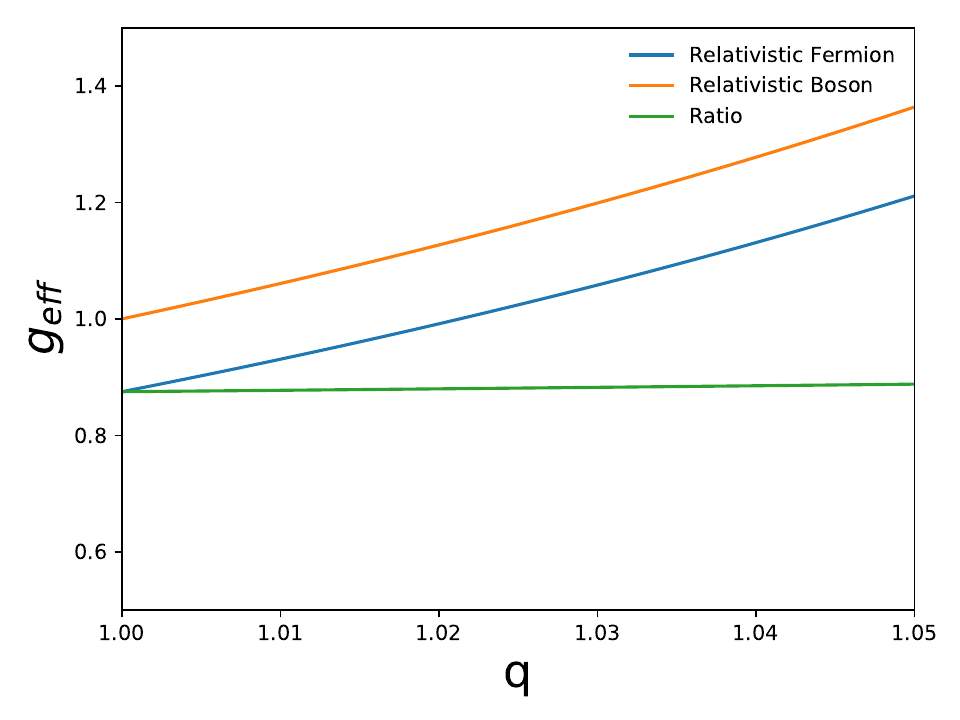}}
\caption{Left panel: The effective number of relativistic degrees of freedom for a massless bosonic (orange) or fermionic (blue) degree of freedom, as a function of $q$. While the number of effective degrees of freedom is enhanced significantly relative to the $q=1$ case as $q$ increases, the ratio of the bosonic case to the fermionic case (green) is nearly constant.} \label{relgeff_fig}
\end{figure}

\begin{figure}
\hspace{-0.4cm}
\centerline{\includegraphics[width=3.2in,angle=0]{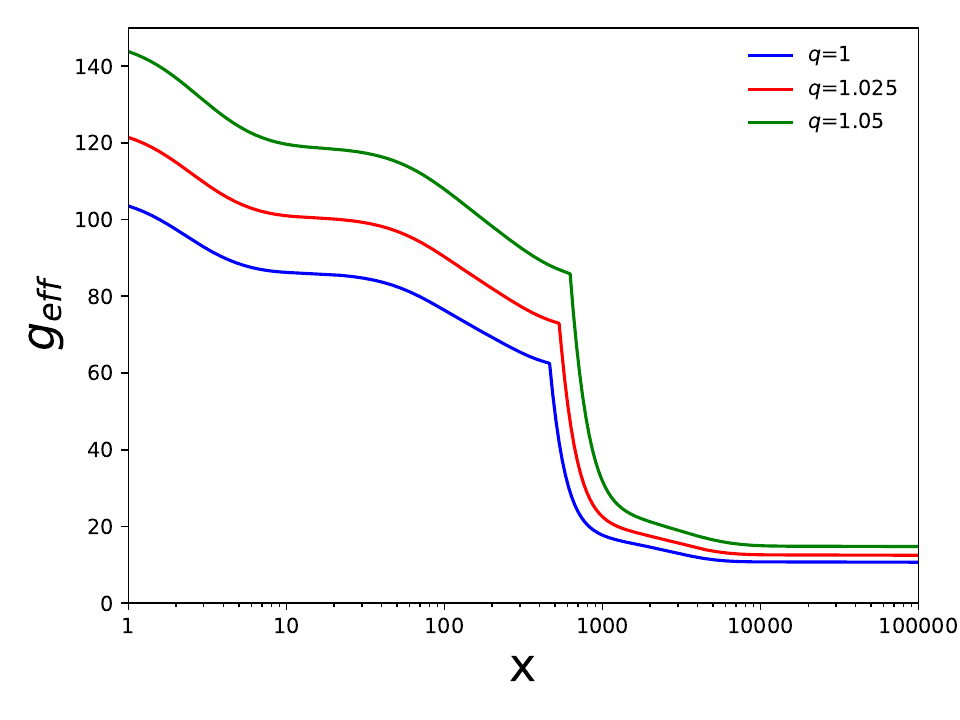}
\includegraphics[width=3.2in,angle=0]{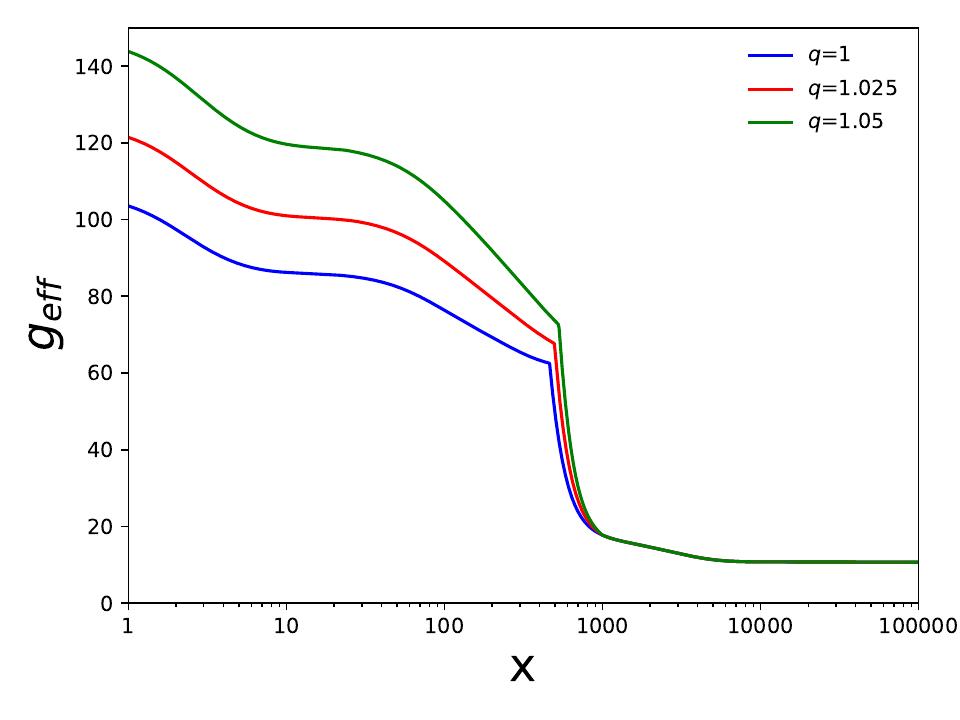}}
\caption{Left panel: The number of relativistic degrees of freedom in the SM, $g_{\textrm{eff}}$ for constant $q=1.05$ (green), $q=1.025$ (red), and $q=1$ (blue) as a function of $x=m/T$, taking $m=100$ GeV. The enhancement of the number of relativistic degrees of freedom slightly changes the temperature of the QCD phase transition, from $T\simeq 216$ MeV for $q=1$ to $T\simeq 160$ MeV for $q=1.05$, which is reflected in the shifting $x$ value of the sharp corner. Right panel: The number of relativistic degrees of freedom in the SM, with $q$ evolving from 1.05 (1.025) at DM freeze-out to 1 as described in section \ref{disc-sec} shown in green (red). } \label{geff_fig}
\end{figure}

The degrees of freedom for a massive species $i$ with $g_i$ spin degrees of freedom can be determined by 

\begin{equation} \label{geff}
    g_{\textrm{eff},i}(T) = \frac{15}{\pi^4} g_i x_i^4 \int_1^{\infty} d\gamma \gamma^2 \sqrt{\gamma^2-1}\left[ \frac{1}{e_q(-x_i \gamma)}+z_i \right]^{-q},
\end{equation}

\begin{equation}
    h_{\textrm{eff},i}(T) = \frac{45}{4 \pi^4} g_i x_i^4 \int_1^{\infty} d\gamma \sqrt{\gamma^2-1} \left( \frac{4\gamma^2-1}{3} \right)\left[ \frac{1}{e_q(-x_i \gamma)}+z_i \right]^{-q},
\end{equation}

\noindent where $x_i = m_i/T$. We note that the above expressions differ somewhat from expressions proposed in the literature which simply replace $e^x \rightarrow e_q(x)$ in the standard $q=1$ definitions \cite{guha2019model}. For a massless species, $g_{\textrm{eff},i}$ can be found by replacing $\gamma^2 \sqrt{\gamma^2-1}$ by $\gamma^3$, taking $x_i = 1$, and integrating from 0 to $\infty$. For massless species $h_{\textrm{eff},i} = g_{\textrm{eff},i}$, as in the $q=1$ case. Fig. \ref{relgeff_fig} shows that as $q$ increases, the effective number of relativistic degrees of freedom (defined relative to the $q=1$ case) increases substantially for both bosons and fermions. The relative number of relativistic degrees of freedom between fermions and bosons for a given $q$, however, changes only very slightly with $q$, going from 7/8 for $q=1$ to 0.888 for $q=1.05$. The total number of relativistic degrees of freedom in the SM is shown as a function of $x$, with $m = 100$ GeV, for various values of $q$ in Fig. \ref{geff_fig}. The temperature of the QCD phase transition is approximated by the temperature at which a gas of mesons and baryons would have lower energy density than a gas of unconfined quarks and gluons, and changes with $q$ from $T_{QCD} \simeq 216$ MeV for $q=1$ to $T_{QCD} \simeq 160$ MeV for $q=1.05$.

Using $H^2 = 8\pi G \rho /3$, and replacing $s$ and $\rho$ by Eq. \ref{rhos}, the Boltzmann equation becomes

\begin{equation} \label{boltzf}
    \frac{dY}{dx} = - \frac{m x^4}{\sqrt{\frac{8 \pi^3 G}{90}}\left( \frac{2 \pi^2}{45} \right)\sqrt{g_{\textrm{eff}}(T)} h_{\textrm{eff}}(T)}\left[ C_{non-eq} - C_{eq} \right].
\end{equation}

\section{Boltzmann Evolution and Results} \label{disc-sec}

To this point we have considered the case of a fixed $q \neq 1$, but the $q$ history of the universe requires some toy model. In order to explore the impact of $q$ on DM freeze-out, we consider a fixed $q_0 \neq 1$ for the SM+DM bath before freeze-out. By the time of the CMB, the photon bath must be consistent with $q=1$ \cite{tsallis1995generalization}, so there must be some transition between the two values of $q$. It is nontrivial to connect two baths of different $q$ values, so for the purposes of our toy model we consider a single $q$ for the universe, which evolves adiabatically from $q=q_0$ up to the point of DM freeze-out to $q=1$ as the universe expands and cools after DM freeze-out. In order to avoid conflicts with cosmological observables, we take $q=1$ at $x=1000$, corresponding to $T \sim 10-1000$ MeV depending on the DM mass. Defining the freeze-out value of $x$ via $Y(x_f) = 2.5 Y_{eq}(x_f)$, we parameterize the $q$ history as:

\begin{equation} \label{qhist}
    q(x) = \left\{
	\begin{array}{ll}
		q_0  & \mbox{if } x \leq x_f \\
		1+(q_0-1)\frac{1000-x}{1000-x_f}& \mbox{if } x_f< x \leq 1000 \\
		1 & \mbox{if } 1000 < x 
	\end{array}	
	\right.
\end{equation} 

For the purposes of this toy model, we consider $q_0 \in [1,1.05]$. We emphasize that this choice of $q$ history is not unique, and the impact of varying $q$ histories merits further study. For example, if $q$ relaxes to 1 much more quickly than in the parameterization of Eq. \ref{qhist}, then we expect the effect of the generalized collision integrals will be greatly reduced (as they will more rapidly approach the usual $n^2$ result) and the main determinant of the target cross section will be that the DM remain in chemical equilibrium until $Y$ is close to $Y_\infty$, as in the $q=1$ case. We note, however, that since values of $q>1$ produce larger equilibrium number densities due to the effects on the phase space distributions seen in Section \ref{stat-sec}, freeze-out would necessarily occur much later in such $q$ histories in order to adequately deplete $Y$ before the DM freezes out. 

For the purpose of numerical study, we take the DM to be a particle with $g_\chi =2$, and consider $m_\chi = 10$, $100$, $1000$ GeV. We consider the case of $s$-wave annihilation, $\sigma v_{lab} \approx a_0$, but generalizing to $p-$ or $d-$wave annihilation by replacing $\sigma v_{lab} = a_1 \epsilon$ or $a_2 \epsilon^2$ in the collision integrals of Eq. \ref{collsv} is conceptually straightforward. In order to numerically solve Eq. \ref{boltzf}, which is stiff, it is useful to rewrite $Y(x) = e^{f(x)}$ and instead solve the differential equation for $f(x)$ \cite{steigman2012precise}:

\begin{equation}
    \frac{df}{dx} = - \frac{m x^4}{\sqrt{\frac{8 \pi^3 G}{90}} \left( \frac{2 \pi^2}{45} \right)\sqrt{g_{\textrm{eff}}(x)} h_{\textrm{eff}}(x)}  e^{-f(x)}\left[ C_{non-eq} - C_{eq} \right].
\end{equation}

We note that since $C_{non-eq}$ depends on $\mu$ and the $q$-exponential cannot be neatly factorized, one must numerically solve for $\mu = \mu(x)$ at every step based on the value of $Y$, and then numerically calculate $C_{non-eq}$ and $C_{eq}$. At $x=1000$ we assume that annihilation stops completely, and thus that $Y_{1000} = Y_{\infty}$, since at this point we have returned to the standard Boltzmann evolution and annihilation beyond $x=1000$ is negligible in the usual case. The relic abundance is then given by

\begin{equation}
    \Omega_\chi h^2 = \frac{m_{\chi} Y_{\infty} s_0 h^2}{\rho_c},
\end{equation}

\noindent where $\rho_c = 3 H_0^2 / (8 \pi G)$, and $s_0$ is the entropy density of the universe at present.

\begin{figure}
\centerline{\includegraphics[width=3.2in,angle=0]{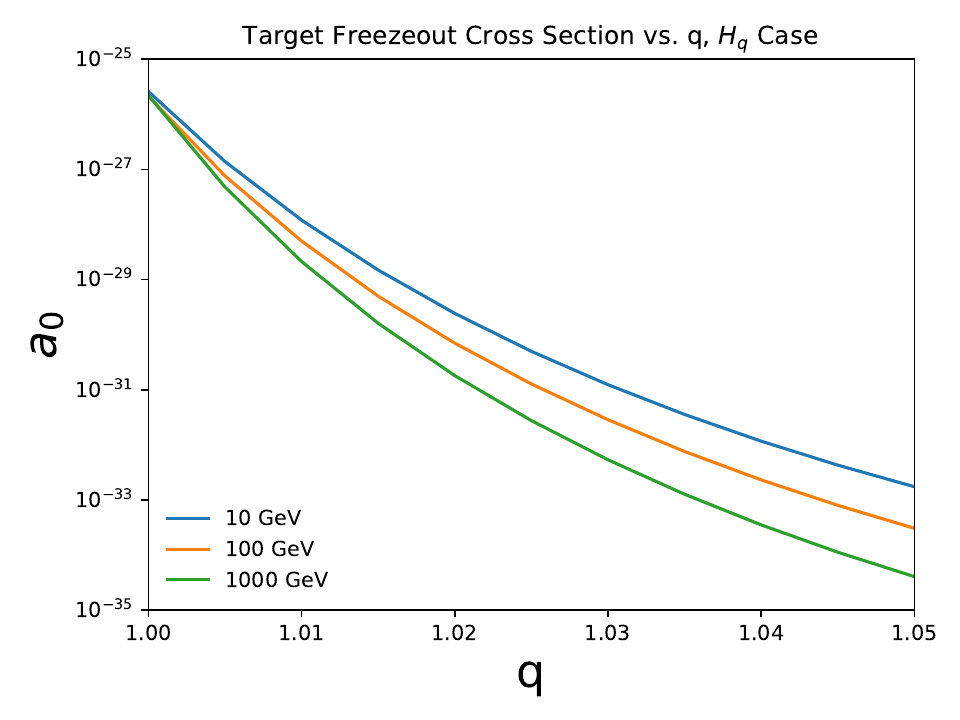}
\includegraphics[width=3.2in,angle=0]{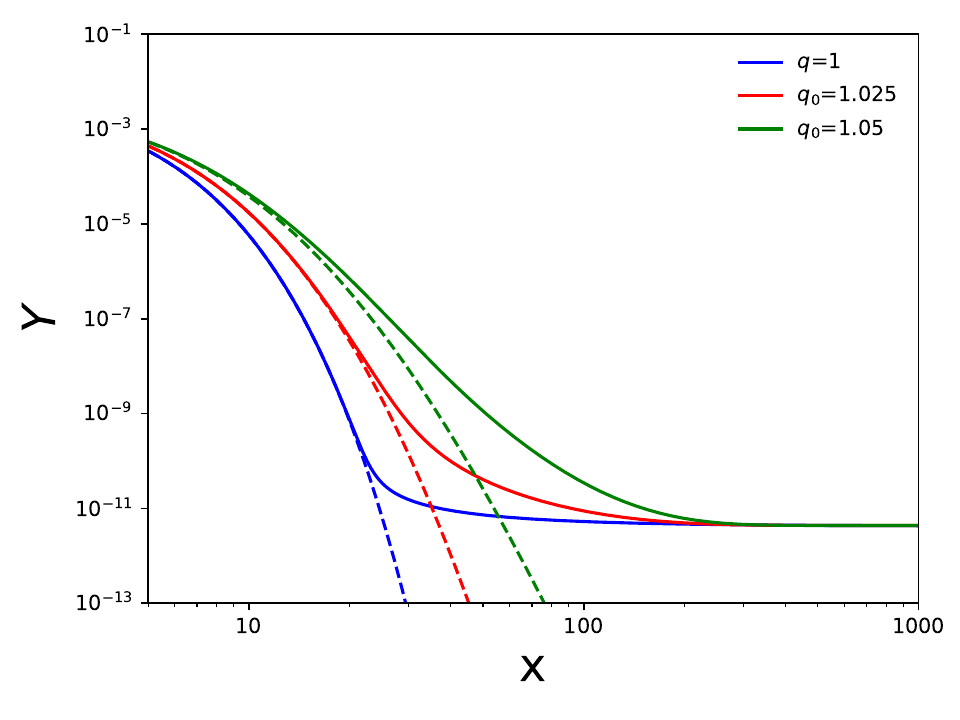}}
\caption{Left: The values of $\sigma v_{lab} = a_0$ which produce the observed relic density of DM as a function of $q_0$ for $m_\chi = 10, 100, 1000$ GeV in blue, orange, and green respectively, with $q$ evolving over time per Eq. \ref{qhist} and the collision integral defined in Eq. \ref{collsv}. For $q=1$ we recover the canonical result $\left< \sigma v \right>\approx 2.2-2.6 \times 10^{-26} \textrm{cm}^3/ \textrm{s}$, depending on the mass of the DM, but as $q$ increases the target value of $a_0$ falls by orders of magnitude, and the spread in target cross sections for this mass range widens to nearly 2 orders of magnitude at $q_0=1.05$. Right: $Y=n/s$ (solid) and $Y_{eq}$ (dashed) as a function of $x=m/T$ for $q=1$ (blue), $q=1.025$ (red), and $q=1.05$ (green), with $m_\chi = 100$ GeV. The generalized collision integrals for $q>1$ cause $Y$ to depart from $Y_{eq}$ much more slowly than in the traditional $q=1$ case and allow DM annihilation to continue significantly after freeze-out, so that for $q=1.05$ we find $x_f=23.7$ even though $Y_{eq}(23.7)$ is nearly 5 orders of magnitude larger than $Y_\infty$. } \label{hq_fig}
\end{figure}

\begin{figure}
\centerline{\includegraphics[scale=0.7]{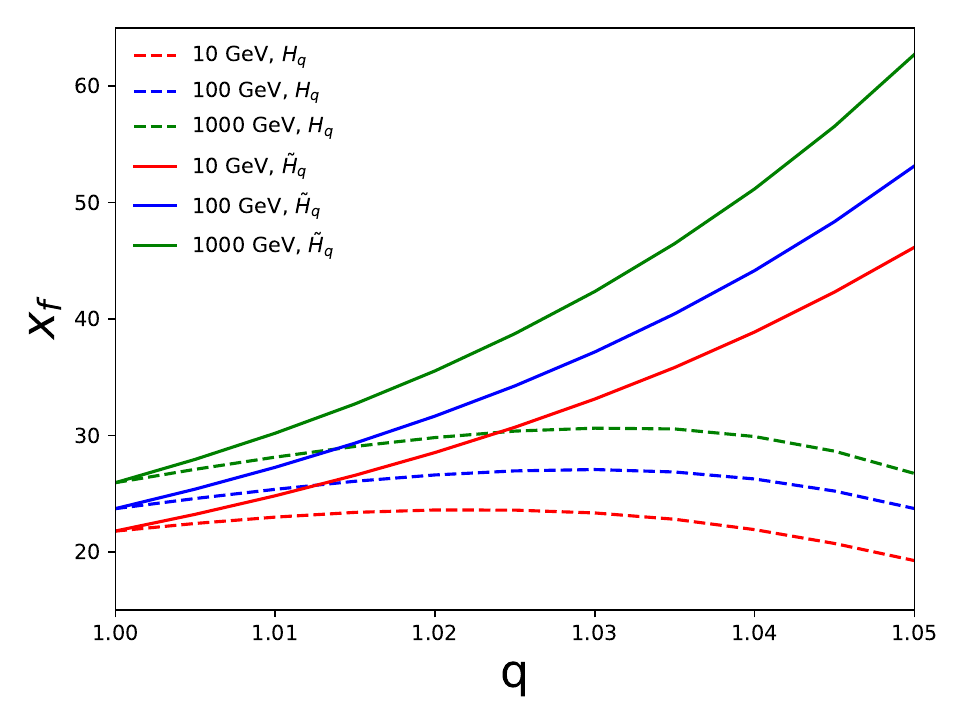}}
\caption{ Values $x_f$, defined as $Y(x_f) = 2.5 Y_{eq}(x_f)$, for the case of $H_q[f_1,f_2] = e_q(\textrm{ln}_q(f_1) + \textrm{ln}_q(f_2) )$ (dashed) and $\tilde{H}_q[f_1,f_2] = e_q(\textrm{ln}_q(f_1) + \textrm{ln}_q(f_2) +(1-q)\textrm{ln}_q(f_1)\textrm{ln}_q(f_2))^q$ (solid). In the first case (dashed), the impact of increasing $q$ is that $H_q$ allows more efficient annihilation after freeze-out, so much so that once $q\gtrsim1.035$ $x_f$ begins to decrease inspite of the fact that a larger $Y_{eq}(x)$ would typically imply later freeze-out. In the second case(solid), the collisional terms reduce to the familiar $n^2$ structure for $s$-wave annihilations, and the larger equilibrium abundance $Y_{eq}(x)$ does require that the DM stay in equilibrium longer to achieve the observed relic density.} \label{xf_fig}
\end{figure}

Fig. \ref{hq_fig} shows the value of $a_0$ which satisfies the observed DM density $\Omega_\chi h^2 = 0.1200 \pm 0.0012$ \cite{aghanim2018planck} as a function of $q_0$ for each of the three masses considered. For $q_0 = 1$ we recover the canonical result $\left< \sigma v \right> = a_0 \approx 2.2-2.6 \times 10^{-26} \textrm{cm}^3/ \textrm{s}$, but even small deviations from $q=1$ have a dramatic impact on the value of $a_0$, which falls by several orders of magnitude as $q$ increases to 1.05. The spread of target $a_0$ for different values of $m_\chi$ increases as well, with a target value of $a_0 \approx 2 \times 10^{-33}  \textrm{cm}^3/ \textrm{s}$ for $m_\chi = 10$ GeV and $a_0 \approx 4\times 10^{-35}  \textrm{cm}^3/ \textrm{s}$ for $m_\chi = 1000$ GeV. Clearly if such small values of $a_0$ yield the observed relic density, we gain a possible understanding of why indirect detection experiments have so far led to null results even with $s$-wave annihilation assumed. For all values of $q$ we find that $x_f \simeq 20-30$, as in the canonical WIMP analysis and shown in Fig. \ref{xf_fig}. This range of $x_f$ implies, however, that there is still significant annihilation occuring beyond freeze-out, as the value of the equilibrium abundance $Y_{eq}$ is several orders of magnitude larger for $q = 1.05$ than for $q=1$ over this range of $x$, as seen in Fig. \ref{hq_fig} right. This Figure shows that additional annihilation after freeze-out lowers $Y$ by the nearly 5 orders of magnitude needed to lower the DM abundance to the observed value. In fact the annihilation after freeze-out is so efficient for larger values of $q \gtrsim 1.035$ that DM begins to freeze out slightly earlier, with freeze-out defined as $Y(x_f) = 2.5 Y_{eq}(x_f)$, which goes against the naive expectation that the enhanced $Y_{eq}(x)$ for larger $q$ values would require DM to stay in equilibrium longer in order to reach the observed relic abundance. The condition that $q = 1$ by $x=1000$ ensures that the annihilation turns off by that time, as we recover the usual Boltzmann equation when $q=1$ and annihilation becomes negligible by $x=1000$ in the canonical case.

\begin{figure}
\centerline{\includegraphics[width=3.2in,angle=0]{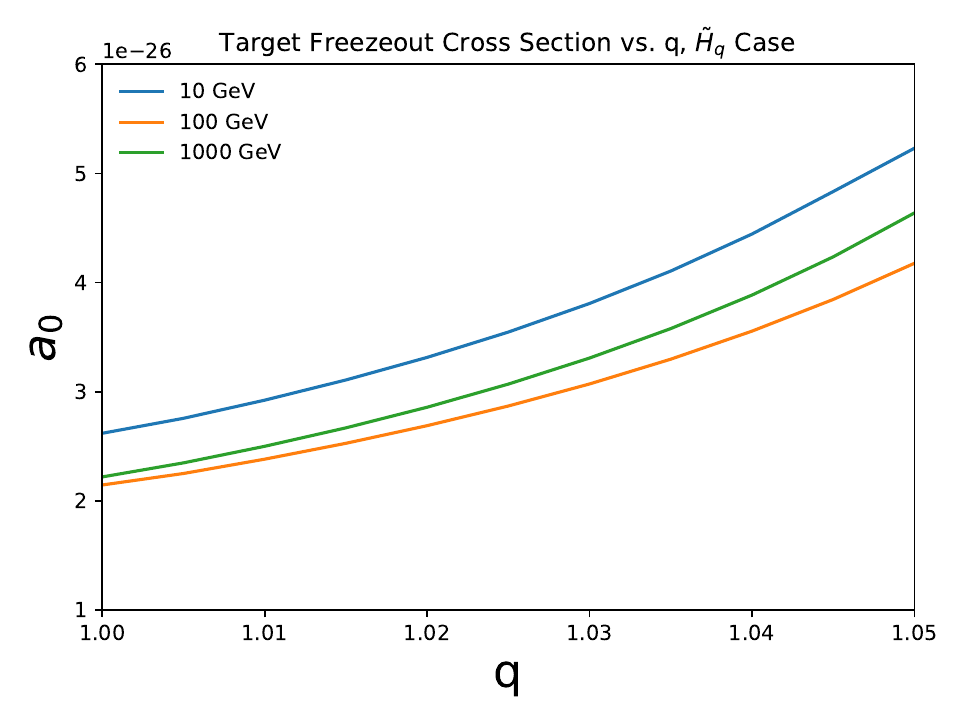}
\includegraphics[width=3.2in,angle=0]{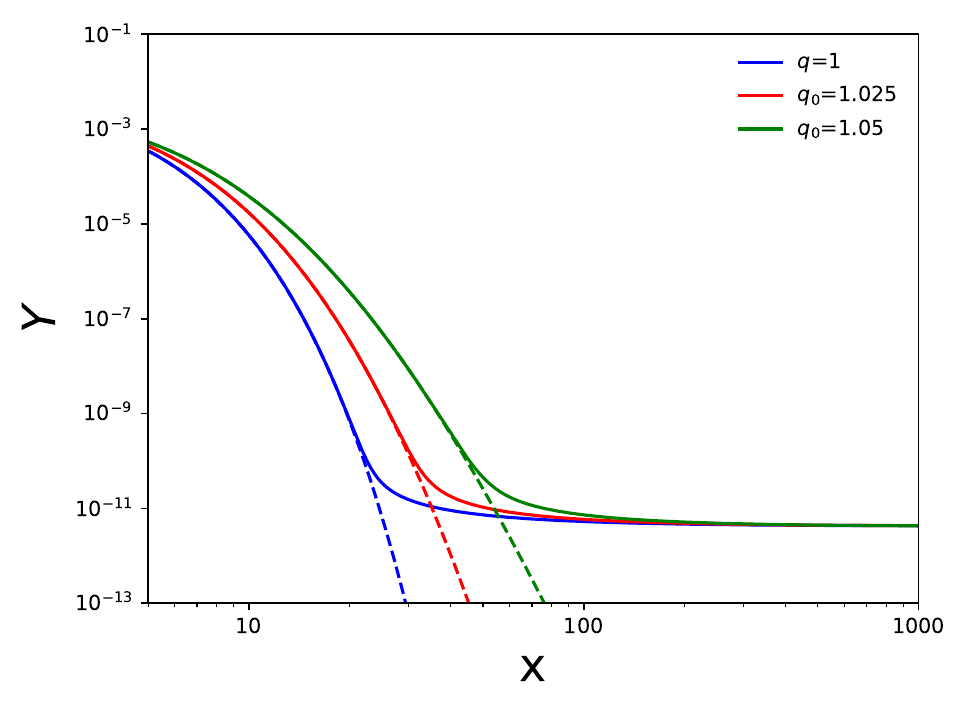}}
\caption{ Left Panel: The values of $\sigma v_{lab} = a_0$ which produce the observed relic density of DM as a function of $q_0$ for $m_\chi = 10, 100, 1000$ GeV in blue, orange, and green respectively, with $q$ evolving over time per Eq. \ref{qhist} and the modified collisional integral described by and below Eq. \ref{fqmod}. Right Panel: $Y=n/s$ (solid) and $Y_{eq}$ (dashed) as a function of $x=m/T$ for $q=1$ (blue), $q=1.025$ (red), and $q=1.05$ (green). For $q> 1$ the equilibrium abundance falls off more slowly than in the $q=1$ case, requiring a slightly larger cross section in order for the DM to stay in equilibrium long enough to reach the observed relic abundance.} \label{fq_fig}
\end{figure}

In Section \ref{boltz-sec}, we determined the functional form of $H_q[f_1,f_2]$ by considering the production of entropy in a scattering process. Naively one might have expected that $H_q$ should be proportional to the mean occupation numbers of species 1 and 2, as in the $q=1$ case. As a toy extension of the work above, we consider a case where the entropy production of a 2 to 2 scattering process has some degree of non-extensivity, so that schematically 

\begin{equation}
    \partial_\mu S^\mu = \int [\textrm{ln}_q(f_1) + \textrm{ln}_q(f_2)+(1-q)\textrm{ln}_q(f_1)\textrm{ln}_q(f_2)](\tilde{H}_q[f_1,f_2]-\tilde{H}_q[f_3,f_4]) + 12 \leftrightarrow 34.
\end{equation} 
In this case, we could consider
\begin{equation} \label{fqmod}
    \tilde{H}_q[f_1,f_2] = e_q(\textrm{ln}_q(f_1) + \textrm{ln}_q(f_2) +(1-q)\textrm{ln}_q(f_1)\textrm{ln}_q(f_2))^q = f_1^q f_2^q,
\end{equation}
 the product of mean occupation numbers. While this model may lead to different equilibrium function distributions due to a differing entropy structure, we assume that the distributions are approximately the same as Eq. \ref{fq} as a toy example. This has the impact of setting $C_{non-eq} = a_0 n^2$ and $C_{eq} = a_0 n_{eq}^2$ in Eq. \ref{boltzf} for the case of $s$-wave annihilation, so that the generalized thermodynamics enters primarily through the altered values of $g_{\textrm{eff}}$ and $h_{\textrm{eff}}$. Fig. \ref{fq_fig} shows the values of $a_0$ which produce the appropriate relic abundance for this case. As seen in the Figure, this extension has a relatively mild impact on the target values of $a_0$ compared to the first case, with the target values of $a_0$ rising by a factor of $\approx 2$ as $q$ increases to 1.05. This is due to the fact that $n_{eq}$ falls more slowly with $x$ as $q$ deviates from 1, so the DM must stay in chemical equilibrium longer in order to reach the observed relic density. Indeed while $x_f \simeq 20-25$ for $q=1$, depending on the DM mass, we find here that $x_f \simeq 46-63$ for $q=1.05$, as shown in the solid lines of Fig. \ref{xf_fig}.

\section{Conclusions}

In this work we examined the effect of a generalized thermodynamics on the WIMP mechanism for DM production in the early universe. We have demonstrated that this generalization requires modification of both the equilibrium phase space distributions of particle species as described in Section \ref{stat-sec}, as well as the collision integral term of the Boltzmann equation as described in Section \ref{boltz-sec}. These modifications make an analytical analysis of DM in the early universe difficult, so we study the Boltzmann equation numerically and search for target cross sections for $s$-wave annihilation which produce the observed DM abundance for $g_\chi=2$, $m_\chi$ = 10, 100, and 1000 GeV and $q_0 \in [1,1.05]$. We find that as $q$ deviates from 1 there is a sharp drop off in the target cross section when considering the case $H_q[f_1,f_2] = e_q(\textrm{ln}_q(f_1)+\textrm{ln}_q(f_2))$, with the target values falling from $2.2-2.6 \times 10^{-26} \textrm{cm}^3/\textrm{s}$ for $q=1$ to $\approx 4\times 10^{-35} - 2\times 10^{-33} \textrm{cm}^3/\textrm{s}$ for $q=1.05$ and $m_\chi$ in the range $10-1000$ GeV. This large drop-off and wide range of possible values for $a_0$ depending on the DM mass motivates a continued program of indirect detection searches at lower values of $\sigma v$, as present experiments are only sensitive to values of $\sigma v$ near the usual $q=1$ limit \cite{oakes2019combined, magic2016limits, ahnen2018indirect, albert2018dark}. The lower target values of $a_0$ for $q>1$ also indicate that DM searches at the LHC may need to aim for smaller cross sections, as a decreased DM annihilation cross section implies a smaller production cross section. We note, however, that there is some model dependence on the exact details of the interactions between the SM and DM when translating $a_0$ into a production cross section at the LHC.

In the future it may be interesting to study the impact of the effective high energy cutoff which occurs for the case $q<1$ on the WIMP scenario. This would have dramatic impacts on $p$- and $d$-wave annihilation, and suppressed values of $n_{eq}$ relative to the $q=1$ case could mean a significantly earlier freeze-out than the typical $x_f \simeq 20-30$ found in this work. It would also be interesting to consider alternative generalizations of thermodynamics, such as the case described in ref \cite{teweldeberhan2005cut}, and their impact on the early universe to study qualitative and quantitative differences in the thermal freeze-out story, as we have seen that even small departures from $q=1$ can greatly impact target values of $\sigma v$. Alternative $q$ histories also merit further study, as it would be interesting to examine quantitatively the robustness of the conclusions presented here under various relaxations of $q\rightarrow1$. Finally, it may be interesting to examine what sort of cosmological signals could arise from $q\neq1$ as further probes of non-standard thermodynamics in the early universe. For instance, the relative effective number of degrees of freedom in relativistic bosons and fermions may impact observables such as $N_{\textrm{eff}}$ , which could be used to constrain models of $q$ evolution.

\section*{Acknowledgements}
This work was supported by the Department of Energy, Contract DE-AC02-76SF00515.

%-----------------------------------------------------------------------------------------------------------------------------------------------------------------------------

\end{document}